\documentclass{elsarticle}

\usepackage[english]{babel}
\usepackage{tabularx}
\usepackage{amsmath, amsfonts, amssymb, bm} 
\usepackage{graphicx}
\usepackage[letterpaper,top=2cm,bottom=2cm,left=3cm,right=3cm,marginparwidth=1.75cm]{geometry}
\usepackage{esvect}
\usepackage{physics}
\usepackage{float}
\usepackage{breqn}

\usepackage{subcaption} 
\usepackage[table]{xcolor}
\usepackage{booktabs}
\usepackage{algpseudocode}
\usepackage{algorithm}
\usepackage[colorlinks=true, allcolors=blue]{hyperref}
\usepackage{tikz}
\usepackage{array}
\usepackage{hhline}
\usepackage{xcolor}
\usepackage{graphicx}
\definecolor{headercolor}{RGB}{63,81,181}
\definecolor{rowcolor1}{RGB}{224,235,255}
\definecolor{rowcolor2}{RGB}{240,240,240}

\usetikzlibrary{shapes.geometric, arrows}
\tikzstyle{startstop} = [rectangle, rounded corners, minimum width=3cm, minimum height=1cm,text centered, draw=black, fill=red!30]
\tikzstyle{io} = [trapezium, trapezium left angle=70, trapezium right angle=110, minimum width=3cm, minimum height=1cm, text centered, draw=black, fill=blue!30]
\tikzstyle{process} = [rectangle, minimum width=3cm, minimum height=1cm, text centered, draw=black, fill=orange!30]
\tikzstyle{decision} = [diamond, minimum width=3cm, minimum height=1cm, text centered, draw=black, fill=green!30]
\tikzstyle{arrow} = [thick,->,>=stealth]

\begin{document}
\begin{frontmatter}

\title{Predicting Crack Nucleation and Propagation in Brittle Materials Using Deep Operator Networks with Diverse Trunk Architectures}

\author[inst1]{Elham Kiyani}
\author[inst2]{Manav Manav}
\author[inst3]{Nikhil Kadivar}
\author[inst2]{Laura De Lorenzis}
\author[inst1]{George Em Karniadakis}

\affiliation[inst1]{organization={Division of Applied Mathematics}, 
                     university={Brown University,}, 
                     addressline={182 George Street}, 
                     city={Providence}, 
                     postcode={02912}, 
                     state={RI}, 
                     country={USA}}

\affiliation[inst2]{organization={Department of Mechanical and Process Engineering}, 
                     university={ETH Zurich,}, 
                     addressline={Tannenstrasse 3}, 
                     city={Zurich}, 
                     postcode={8092}, 
                     country={Switzerland}}

\affiliation[inst3]{organization={School of Engineering}, 
                     university={Brown University,}, 
                     addressline={184 Hope Street}, 
                     city={Providence}, 
                     postcode={02912}, 
                     state={RI}, 
                     country={USA}}

\begin{abstract}
Phase-field modeling reformulates fracture problems as energy minimization problems and enables a comprehensive characterization of the fracture process, including crack nucleation, propagation, merging and branching, without relying on ad-hoc assumptions. 
However, the numerical solution of phase-field fracture problems is characterized by a high computational cost. 
To address this challenge, in this paper, we employ a deep neural operator (DeepONet) consisting of a branch network and a trunk network to solve brittle fracture problems. We explore three distinct approaches that vary in their trunk network configurations. In the first approach, we demonstrate the effectiveness of a two-step DeepONet, which results in a simplification of the learning task.
In the second approach, we employ a physics-informed DeepONet, whereby the mathematical expression of the energy is integrated into the trunk network’s loss to enforce physical consistency. The integration of physics also results in a substantially smaller data size needed for training.
In the third approach, we replace the neural network in the trunk with a 
 Kolmogorov-Arnold Network and train it without the physics loss. Using these methods, we model crack nucleation in a one-dimensional homogeneous bar under prescribed end displacements, as well as crack propagation and branching in single edge-notched specimens with varying notch lengths subjected to tensile and shear loading. We show that the networks predict the solution fields accurately and the error in the predicted fields is localized near the crack.
\end{abstract}
\end{frontmatter}

\section{Introduction}

The accurate prediction of fracture phenomena in brittle materials is of crucial importance in practical applications. 
Griffith's seminal work in 1920~\cite{griffith1921vi} laid the groundwork for brittle fracture theory; according to Griffith, crack propagation is dictated by the balance between the consumption of energy associated to the creation of new surfaces and the release of bulk elastic energy as the crack propagates. More recently, Griffith's criterion was recast in a variational form~\cite{francfort1998revisiting} giving rise to a free discontinuity problem. Regularization of this problem led to the variational phase-field approach to fracture~\cite{bourdin2000numerical, bourdin2008variational}, which gained enormous popularity over the past two decades ~\cite{ambati2015review} and became an increasingly powerful framework for predicting a variety of fracture problems including thermal ~\cite{bourdin2014morphogenesis}, drying-induced ~\cite{luo2023phase}, hydraulic ~\cite{heider2021review} and ductile fracture~\cite{alessi2018comparison}, among many others. The approach introduces a continuous damage or phase-field variable varying between 0, representing intact conditions, and 1, representing fully damaged conditions. The displacement and phase fields are governed by a system of non-linear coupled partial differential equations (PDEs), which arise from the stationarity conditions of the total energy functional, i.e. as necessary conditions for the system to reach a state of minimum energy. This model leads to unprecedented flexibility in the simulation of complex crack behaviors, such as nucleation, branching, and merging, with no need for explicit crack tracking, thereby greatly simplifying implementation. However, it introduces a small length scale which is computationally expensive to resolve in a discretized setting. Moreover, non-convexity of the energy functional requires special attention in the numerics~\cite{lorenzis2020numerical}.


Machine learning recently emerged as a powerful set of tools for modeling of complex material behavior including fracture, see the recent review in \cite{fuhg2024review}.
In particular, physics-informed neural networks (PINNs)~\cite{raissi2019physics}, a versatile framework for solving forward and inverse problems~\cite{haghighat2022physics,rasht2022physics, kiyani2024characterization, kiyani2023framework}, were also explored in the context of fracture modeling within the phase-field framework~\cite{goswami2020transfer, manav2024phase}, as well as for damage modeling in quasi-brittle materials~\cite{zheng2022physics}, and for modeling of fatigue crack growth~\cite{chen2024crack}.
Even more recently, operator learning  emerged as a promising alternative for solving PDEs, particularly in problem settings with changing boundary or initial conditions, or source functions. Operator learning focuses on mapping infinite-dimensional input and output spaces, often implemented by discretizing them onto finite grids for practical computation
~\cite{lu2019deeponet,shih2024transformers,kovachki2023neural,shukla2024deep,oommen2024integrating,CiCP-35-1194}. A notable example is the Deep Operator Network (DeepONet)~\cite{lu2019deeponet}, which uses two subnetworks: the branch network to encode input functions and the trunk network to encode spatial or temporal coordinates. This architecture allows for efficient learning of complex operators that arise in scientific and engineering applications. In addition to traditional data-driven DeepONets, physics-informed DeepONets combine the strengths of data-driven and physics-based approaches by embedding PDE residuals directly into the loss function, thereby ensuring that the learned operators respect physical laws~\cite{wang2021learning}. This fusion enhances predictive accuracy with limited data and enables efficient solutions for parametric PDE problems, making physics-informed DeepONets a powerful tool for solving complex, data-constrained problems. 
An optimized two-step training strategy which decouples the optimization of the branch net and the trunk net was also introduced to address challenges in simultaneous optimization, leading to improved accuracy and generalization across diverse applications~\cite{lee2024training, peyvan2024riemannonets}.

Thus far, only a few studies explored DeepONets for phase-field modeling and/or fracture problems. A physics-informed  DeepONet was proposed to enable real-time predictions of pattern-forming systems governed by gradient flows, exemplified by phase-field models such as the Allen–Cahn and Cahn–Hilliard equations ~\cite{li2023phase}, but was not applied to fracture problems. A DeepONet  framework informed by  the variational formulation for 
quasi-brittle fracture
was proposed in~\cite{GOSWAMI2022114587}. While the authors demonstrated that this method could predict crack paths and calculate displacement and damage at each displacement value  from the previous step, their proposed approach was generally inefficient due to error propagation throughout the predictions.

In this study, we investigate a two-step DeepONet~\cite{lee2024training, peyvan2024riemannonets} approach to model brittle fracture, using both a data-driven and a physics-informed approach, with the data and (for the latter) the physics  relying on the phase-field approach to fracture. 
We start with a data-driven DeepONet, in which both the branch net and the trunk net are implemented as traditional multilayer perceptrons (MLPs). The branch net captures the effect of changing boundary conditions or varying initial notch sizes, while the trunk net focuses on the spatial coordinates. We then extend this framework to a physics-informed two-step DeepONet, whereby the energy functional governing phase-field fracture is integrated into the model. This modification allows us to minimize the energy directly as part of the trunk net loss, leading to more physically consistent predictions and a substantially small data size needed for training. Finally, to enhance the model's accuracy and adaptability, we employ a modified version of DeepOKAN~\cite{shukla2024comprehensive} which integrates Kolmogorov-Arnold Networks (KANs)~\cite{liu2024kan} into the vanilla DeepONet architecture. 
Inspired by Kolmogorov Networks~\cite{sprecher2002space, koppen2002training}, KANs dynamically adjust activation patterns in response to the input data, improving both accuracy and interpretability. 

In our setting, we use a KAN only in the trunk net.
One of the key contributions of this work lies in the use of a two-step DeepONet for modeling crack nucleation, propagation and branching, which enhances the accuracy of the prediction of the solution fields. Unlike in previous approaches, the pre-training of the trunk network in our method plays a crucial role, as suggested by the resemblance of the basis functions of the trunk network output to the solution fields, offering valuable insights into the outcomes of the complete training process. This pre-training significantly enhances prediction accuracy while streamlining the overall training process. Furthermore, our approach enables the prediction of the displacement and phase fields directly from the initial crack size at each prescribed boundary displacement value, overcoming the problem of error accumulation. Despite the challenges associated with the accurate prediction of the solution fields in phase-field modeling, we successfully trained the two-step DeepONet in a data-driven framework using only 45 samples, and in the physics-informed framework using only 10 samples.

The paper is structured as follows: Section~\ref{sec:Phase-field-fracture-model} overviews the phase-field brittle fracture model used in this work. Section~\ref{sec:Operator_Learning} introduces the data-driven DeepONet and its two-step training process together with the details of our network design and training approach. In Section~\ref{sec:Data-Driven}, we present an  analysis of the two-step DeepONet for data-driven crack nucleation and propagation prediction. Section~\ref{sec:Physics informed-Two-stepDeepont} details the integration of physics-informed principles into the trunk net loss function and showcases the results of the physics-informed two-step DeepONet. In Section~\ref{sec:KAN}, we present the DeepOKAN architecture and its results. Finally, in Section 
\ref{sec:Comparison} we discuss the comparative performance of the three methods, and we conclude the paper with a summary in Section
\ref{sec:Summary}.


\section{Phase-field modeling of brittle fracture}\label{sec:Phase-field-fracture-model}

In this section, we briefly describe the phase-field fracture model for isotropic brittle materials under quasi-static loading which we adopt in this work.
Let $\Omega \subset \mathrm{R}^d$, $d \in {1, 2, 3}$ be an open bounded domain denoting an arbitrary body with external boundary $\Gamma$, possibly containing a crack set $\Gamma_c$ (see Figure 1(a)). We denote the Dirichlet and Neumann portions of the boundary as  $\Gamma_D$ and $\Gamma_N$, respectively, such that $\Gamma_D \cup \Gamma_N = \Gamma$ and $\Gamma_D \cap \Gamma_N = \emptyset$. $\Gamma_D$ in turn generally consists of a homogeneous Dirichlet boundary, $\Gamma_{D,0}$, and a nonhomogeneous Dirichlet boundary, $\Gamma_{D,1}$. $\Gamma_{N,1}$ denotes the nonhomogeneous Neumann boundary. The displacement of a material point $\bm{x}$ in the body is denoted by $\bm{u}(\bm{x})$, and we assume infinitesimal deformations so that the strain tensor is given by $\bm{\varepsilon} = \mathrm{sym}(\nabla \bm{u})$. The material is linear elastic and isotropic, with the strain energy density function given by
\[
\Psi(\bm{\varepsilon}) = \frac{1}{2} \lambda \, \text{tr}^{2}(\bm{\varepsilon}) + \mu \, \bm{\varepsilon} \cdot \bm{\varepsilon},
\]
where $\lambda$ and $\mu$ are the Lamé constants. 
\begin{figure}[!tbh]
\centering
\includegraphics[width=0.7\textwidth]{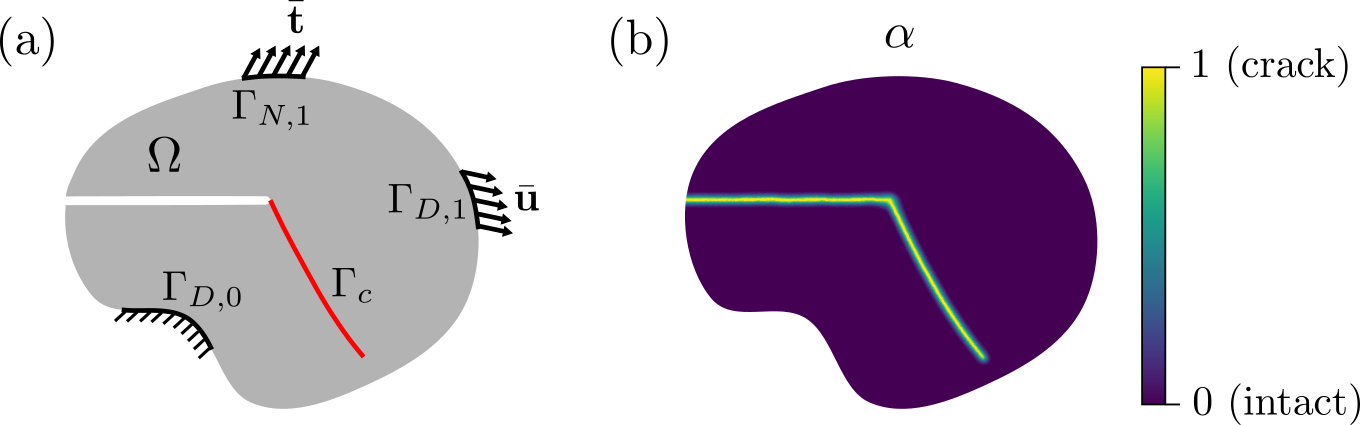}
\caption{(a) A solid body containing a crack-like notch and a sharp crack, and (b) the plot of the phase field $\alpha$ regularizing them.}
\label{fig:page16}
\end{figure} 
In the phase-field fracture model, the energy functional for a body subjected to surface traction $\bar{\mathbf{t}}$ and body force $\bar{\mathbf{b}}$ is given by:
\begin{equation}\label{eq:2}
    \begin{aligned}
        \mathcal{E}(\bm{u},\alpha)
        = &\underbrace{\int_{\Omega} (g(\alpha)  \Psi^{+}(\bm{\varepsilon}(\bm{u})) + \Psi^{-}(\bm{\varepsilon}(\bm{u}))) \, d\bm{x}}_{\mathcal{E}_{el}(\bm{u},\alpha)} + \underbrace{\frac{G_c}{c_w} \int_{\Omega} \left(\frac{w(\alpha)}{l} + l \lvert \nabla \alpha \rvert ^2\right) \, d\bm{x}}_{ \mathcal{E}_{d}(\alpha)} \\
        &- \int_{\Omega} \bar{\mathbf{b}} \cdot \bm{u} \, d\bm{x} - \int_{\Gamma_{N,1}} \bar{\mathbf{t}} \cdot \bm{u} \, ds,
    \end{aligned}
\end{equation}
where $\mathcal{E}_{el}$ and $\mathcal{E}_{d}$ are the elastic and damage energies, respectively, and $\alpha \in [0,~1]$ denotes the phase field. Specifically, $\alpha =0$ corresponds to no damage, while $\alpha =1 $ denotes complete damage, and values in between represent increasing levels of damage with increasing $\alpha$. The parameter $l$, satisfying 
$0 < l \ll \mathrm{diam}(\Omega)$,
is the regularization length, which controls the thickness of the transition zone between fully damaged and undamaged material. The degradation
function $g(\alpha)$ governs the reduction of the elastic energy with increasing damage,  while the local dissipation function $w(\alpha)$ represents the energy dissipated through homogeneous damage within a unit volume of the material. The elastic strain energy density $ \Psi$ is decomposed into crack-driving component $ \Psi^{+}$ and crack-resisting component $ \Psi^{-}$, whereby only the former is coupled to the phase field and thus drives damage evolution. The material parameter $G_c$ denotes the critical energy release rate, assumed to be a material property according to Griffith. The constant $c_w$ is a normalization factor; it ensures that the phase-field regularized fracture energy converges to the fracture energy of the Griffith model as $l$ approaches zero. In this work, we adopt the following forms of the degradation and local dissipation functions, which are widely used in the literature:

\begin{align}\label{AT1,AT2-model}
g(\alpha)&=(1-\alpha)^2,\\
w(\alpha)&= \begin{cases}
            \alpha, & \text{AT1 model}\\
            \alpha^2, & \text{AT2 model.}
\end{cases}
\end{align}
Here AT stands for ``Ambrosio-Tortorelli'' model, and the number refers to the exponent of $\alpha$ in the related model. The properties of these models are extensively analyzed in \cite{pham2011}.

Of the multiple approaches for decomposition of the strain energy density discussed in the literature~\cite{amor2009regularized, freddi2009variational, miehe2010thermodynamically, vicentini2024energy}, we choose the widely used volumetric-deviatoric split~\cite{amor2009regularized}.
\begin{align}
	&\Psi^+=\frac{1}{2}K\langle\rm tr(\bm{\varepsilon})\rangle_+^2+\mu\mathbf{e}\cdot\mathbf{e}, \nonumber \\
	& \Psi^-=\frac{1}{2}K\langle\rm tr(\bm{\varepsilon})\rangle_-^2,
\end{align}
where $\langle a \rangle_+ = {\rm max}\{0, a\}$ and $\langle a \rangle_- = {\rm min}\{0, a\}$. Also, we introduce the bulk modulus $K=\lambda + \frac{2}{d}\mu$ and the deviatoric strain tensor $\mathbf{e}=\bm{\varepsilon}-\frac{tr(\bm{\varepsilon})}{d}\mathbf{I}$, where $\mathbf{I}$ is the second-order identity tensor.

In a time-discrete quasi-static evolution, the state of the system at the $n^{th}$ loading step is obtained by solving the following minimization problem:
\begin{center}$\operatorname{argmin}_{\bm{u},\alpha} \left\{ \mathcal{E}_n(\bm{u}, \alpha) :  \bm{u} \in V_{\bar{u}_n},~\alpha \in D_{\alpha_{n-1}} \right\},$
\end{center}
where
\begin{equation}\label{eq:energy2}
    \mathcal{E}_n(\bm{u}, \alpha) = \mathcal{E}_{el}(\bm{u}, \alpha) + \mathcal{E}_d(\alpha) - \int_{\Omega} \bar{\mathbf{b}}_n \cdot \bm{u} \, d\bm{x} - \int_{\Gamma_{N,1}} \bar{\mathbf{t}}_n \cdot \bm{u} \, ds,
\end{equation}

\begin{equation}
    V_{\bar{u}_n} = \left\{\bm{u} \in (H^1(\Omega))^d :  \bm{u} = \bm{0} \text{ on } \Gamma_{D,0}, ~\bm{u} = \bar{\bm{u}}_n \text{ on } \Gamma_{D,1}\right\},
\end{equation}

\begin{equation}
    D_{\alpha_{n-1}} = \left\{ \alpha \in H^1(\Omega) :  \alpha \geq \alpha_{n-1} \text{ in } \Omega \right\},
\end{equation}
and $\alpha_{n-1}$ is the phase field from the previous loading step. $V_{\bar{u}_n}$ denotes the set of kinematically admissible displacement fields and $D_{\alpha_{n-1}}$ denotes the set of admissible phase fields satisfying  irreversibility of damage (no healing). In this work, irreversibility of the phase-field variable is enforced by adding the following energetic penalty~\cite{gerasimov2019penalization} to the energy functional in \eqref{eq:2}:

\begin{equation}
	\mathcal{E}_{ir}(\alpha)=\frac{1}{2}\gamma_{ir}\langle \alpha-\alpha_{n-1}\rangle_-^2, \label{eq6a}
\end{equation}
where $\gamma_{ir}$ is the penalty parameter given by
\begin{align}
	\gamma_{ir} = \begin{cases}
            \frac{G_c}{l}\frac{27}{64\mathtt{TOL^2_{ir}}}, & \text{AT1 model}\\
            \frac{G_c}{l}\left(\frac{1}{\mathtt{TOL^2_{ir}}}-1\right), & \text{AT2 model}.
        \end{cases} \label{eq6b}
\end{align}
Here $0<\mathtt{TOL_{ir}}\le 1$ is a prescribed tolerance threshold and is set to  $5\times 10^{-3}$ in the physics-informed training of DeepONets.


\section{Operator learning through DeepONets}\label{sec:Operator_Learning}

DeepONet is a specialized neural network framework engineered to approximate operators — i.e., mappings between infinite-dimensional function spaces — that are crucial for solving parametric PDEs ~\cite{lu2019deeponet}. The input to a DeepONet can be initial conditions, boundary conditions, source functions etc. and it maps an input function, denoted here by \( f\in F \), to the corresponding PDE solution \(g\in G \), where \(F \) and \(G \) are appropriately defined function spaces~\cite{lu2019deeponet},  by approximating the operator \( \mathcal{G}: F\mapsto G \). The DeepONet architecture consists of two distinct subnetworks:

\textbf{Branch Network:}  
This subnetwork encodes the input functions. To encode a function \( f \), the function is first sampled at a finite set of sensor points \( \{\bm{x}_1, \bm{x}_2, \ldots, \bm{x}_k\} \). 
The function values,
\( (f(\bm{x}_1), f(\bm{x}_2), \ldots, f(\bm{x}_k)) \),
are input into the branch network, which transforms them into a latent representation \( \bm{b} \in \mathbb{R}^p \), where \( p \) represents the latent dimension. For a set of input functions \(\{f_1, f_2, \ldots, f_m\}\), the output of the network is \( B \in \mathbb{R}^{m \times p} \) where the \(i^{th}\) row of \(B\) is the latent representation of \(f_i\).

\textbf{Trunk Network:}  
This subnetwork encodes the spatial coordinates of points at which the output of the operator is to be evaluated. Specifically, it takes a spatial coordinate \( \bm{x} \) as input and generates a coordinate-dependent basis  \(\boldsymbol{\phi} \in \mathbb{R}^p\). For a set of spatial coordinates \( \{\bm{x}_1, \bm{x}_2, \ldots, \bm{x}_n\} \), the trunk network generates an output \( \Phi \in \mathbb{R}^{n \times p} \) where the \(j^{th}\) row of \(\Phi\) is the basis corresponding to \(\bm{x}_j\).

The output of the trunk network combines with the output of the branch network to approximate the output of the operator as follows:

\begin{equation}
\mathcal{G}(f_i)(\bm{x}_j)  \approx B_{il}(f_i; {\theta}_B) \Phi_{jl}(\bm{x}_j; {\theta}_T), 
\label{eq:deeponet_outputs7}
\end{equation}
where \( {\theta}_B \) and \( {\theta}_T \) denote the trainable parameters of the branch and trunk networks, respectively, and Einstein summation notation is employed. 

\begin{figure}[!tbh]
    \centering
    \includegraphics[width=0.99\textwidth]{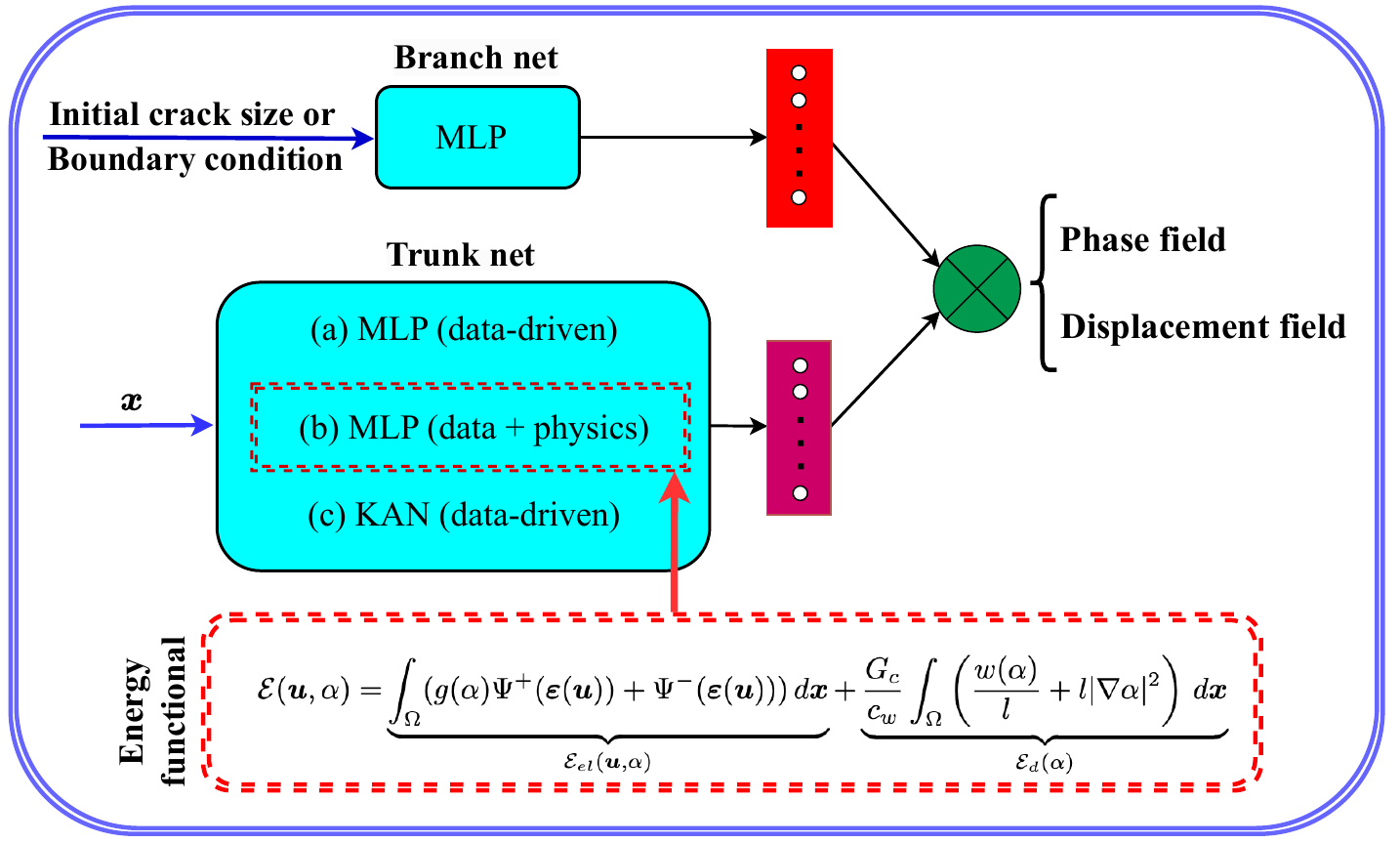}
    \caption{Schematic representation of the DeepONet structure with three different types of trunk networks. The branch network processes the boundary condition or the initial crack size, while the trunk network processes the spatial coordinate \(\bm{x}\) to predict the phase field \(\alpha(\bm{x})\) and the displacement field \(\bm{u}(\bm{x})\).
    }
\label{fig:Deeponet_2}
\end{figure}

The DeepONet framework supports a variety of architectures for the branch and trunk networks. 
In the original formulation, both the branch and trunk networks were implemented as MLPs. 
The DeepOKAN architecture enhances the operator learning framework by incorporating KANs, replacing traditional MLPs with possibly more flexible and adaptive function approximators.
KANs are based on the Kolmogorov-Arnold representation theorem, which provides a rigorous mathematical foundation for approximating multivariate continuous functions. This theorem states that any multivariate function \( f(\bm{x}) \), where \( \bm{x} \) is a point in a bounded domain, can be represented as a finite composition of univariate continuous functions.
A generalized KAN architecture is described by the sequence of layer sizes \( [n_1, \dots, n_{K+1}] \), where \( K \) represents the number of layers in the network. In the context of deeper KANs, the \( j^{th} \) layer can be represented by a matrix \( \Xi_j\), where each element \( \xi_{ij}(\cdot) \) in layer \( j\) is a learnable univariate function mapping inputs to outputs. Let \( N_{\text{in}} \) denote the number of inputs and \( N_{\text{out}} \) denote the number of outputs for a given layer. The original Kolmogorov-Arnold theorem can be visualized as a two-layer KAN: the inner layer has \( N_{\text{in}} = n \) inputs and \( N_{\text{out}} = 2n + 1 \) outputs, while the outer layer combines these outputs to yield a single value, \( N_{\text{out}} = 1 \). The function computed by a deeper KAN is the composition of these layers, expressed as:

\begin{equation}
y = \text{KAN}(\bm{x}) = ({\Xi}_K \circ {\Xi}_{K-1} \circ \dots \circ {\Xi}_1)(\bm{x}).
\end{equation}

Unlike traditional MLPs, which use fixed activation functions at each neuron, KANs incorporate adaptive activation functions on the edges. 
In the original paper by Liu et al.~\cite{liu2024kan}, splines were introduced as replacements for conventional linear weight matrices. The spectral bias phenomenon in KANs was further investigated by Wang et al.~\cite{wang2024expressiveness}, where the authors demonstrated that a single KAN layer does not exhibit spectral bias. Additionally, KANs were shown to have a significantly reduced spectral bias compared to MLPs.

Since all operations in KANs are differentiable, they can be efficiently trained using backpropagation. This formulation extends the Kolmogorov-Arnold theorem to modern deep learning frameworks, providing a promising alternative to MLPs by offering enhanced interpretability. In this work, we use an MLP as a branch network, and a KAN as a trunk network (Figure~\ref{fig:Deeponet_2}).

\subsection{Two-step DeepONet training method}\label{Two-step}
The standard DeepONet framework approximates an operator by simultaneously training both the branch and trunk networks. 
Training both the trunk and branch networks simultaneously typically requires solving a complex optimization problem in a high-dimensional space, which is both non-convex and non-linear, making the training process challenging. To address this issue, a two-step training method has been proposed~\cite{lee2024training}, which breaks down the entire optimization problem into two simpler subproblems.

\begin{figure}[!tbh]
    \centering
    \includegraphics[width=0.89\textwidth]{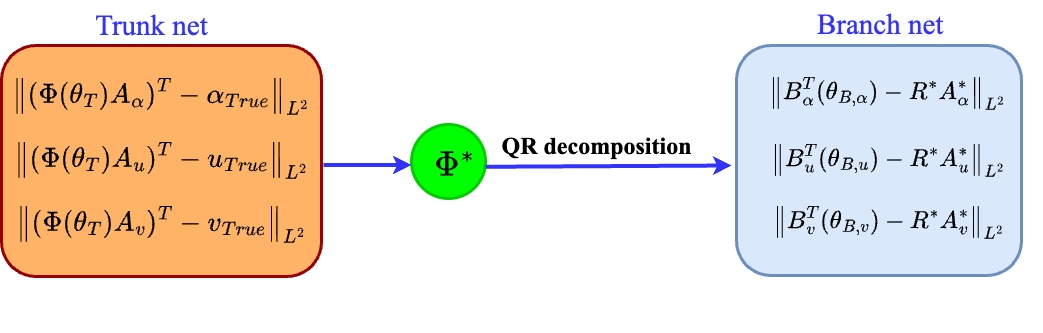}
\caption{Schematic of the two-step DeepONet for predicting the phase field \(\alpha\) and the components \(u\) and \(v\) of the displacement field $\bm{u}$. 
During the training, first, the trunk network parameters, along with matrices \({A}_{u}\), \({A}_{v}\), and \({A}_{\alpha}\), are optimized, followed by the QR factorization of \({\Phi}^*\), and finally the training of the branch network.
}
    \label{fig:Two_step_uvalpha}
\end{figure}

In this section, we provide a brief review of two-step DeepONet, following the framework established in~\cite{lee2024training}. The first step involves learning a set of basis functions through the trunk network. 
Unlike in traditional DeepONet training, the branch network is not used in the first step. Instead, the model optimizes a coefficient matrix \( A \) that captures the relationship between the basis functions and the output. The optimization problem is formulated as:

\begin{equation}
\min_{\theta_T, A} L(\theta_T, A) := \| \Phi(\theta_T) A - U \|.
\end{equation}
where \( \Phi(\theta_T) \in \mathbb{R}^{n \times p} \) represents the trunk network output, which serves as a set of learned basis functions.
 \( A \in \mathbb{R}^{p \times m} \) is a trainable coefficient matrix that maps the learned basis functions to the target outputs.
 \( U \in \mathbb{R}^{n \times m} \) represents the ground truth solution.
The norm \( \| \cdot \| \) can be the \( L_2 \) norm, the Frobenius norm, or any other appropriate norm for measuring the reconstruction error. As we aim to predict both the displacement and the damage fields,
the sequential two-step training approach (see Figure~\ref{fig:Two_step_uvalpha}) begins with the training of the trunk network by minimizing the following loss function with respect to \({\theta}_T \), \({A}_{\alpha}\), \({A}_{u}\), and \({A}_{v}\) in a two-dimensional problem:
\begin{equation}\label{eq:loss_trunk}
\begin{aligned}
\mathcal{L}_{T_{\text{data}}}({\theta}_T, {A}_{\alpha}, {A}_{u}, {A}_{v}) &= 
\lambda_{\alpha} \, \texttt{loss}_{\alpha} + \lambda_{u} \, \texttt{loss}_{u} + \lambda_{v} \, \texttt{loss}_{v} \\
&= \lambda_{\alpha} \left\| {\alpha}_{\theta} - {\alpha}_{\text{True}} \right\|_{L^2} +
\lambda_{u} \left\| {u}_{\theta} - {u}_{\text{True}} \right\|_{L^2} +
\lambda_{v} \left\| {v}_{\theta} - {v}_{\text{True}} \right\|_{L^2}, \\
{\alpha}_{\theta} &= ({\Phi}({\theta}_T) {A}_{\alpha})^T, \quad 
{u}_{\theta} = ({\Phi}({\theta}_T) {A}_{u})^T, \quad 
{v}_{\theta} = ({\Phi}({\theta}_T) {A}_{v})^T.
\end{aligned}
\end{equation}
where \({\Phi}({\theta}_T)\) is the output of the trunk network, and \({A}_{\alpha}, {A}_{u},\) and \({A}_{v} \in \mathbb{R}^{p \times m}\) are trainable matrices.
The arrays \( \alpha_{\theta} \), \( u_{\theta} \), and \( v_{\theta} \) represent the intermediate field predictions generated solely by the trunk network. Correspondingly, the arrays \( \alpha_{\text{True}} \), \( u_{\text{True}} \), and \( v_{\text{True}} \) are \( m \times n \) arrays, containing the ground truth data for damage and displacement fields.

In addition, \(\lambda_{\alpha}\), \(\lambda_{u}\), and \(\lambda_{v}\) are trainable weights that are adjusted during the training process, refer to~\ref{sec:Adaptive weighing}.
After optimization, let \({\theta}_T^*, {A}_{\alpha}^*, {A}_{u}^*,\) and \({A}_{v}^*\) denote the optimized parameters and matrices, where \({\Phi}({\theta}_T^*)\) is assumed to be full rank. To prepare for the next step, either QR decomposition or singular value decomposition (SVD) of \({\Phi}({\theta}_T^*)\) must be performed. In this study, we utilize the QR decomposition:


\begin{equation}
{Q}^* {R}^* = {\Phi}({\theta}_T^*).
\end{equation}

In the second step of training, the branch network is trained to match the projected matrices \( {R}^* {A}_{\alpha}^* \), \( {R}^* {A}_{u}^* \), and \( {R}^* {A}_{v}^* \) corresponding to the three fields. Note that one can choose the configuration of the branch network, and in this paper we employ three independent networks with trainable parameters \({\theta}_{B, \alpha} \), \( {\theta}_{B, u} \), and \( {\theta}_{B, v} \) corresponding to \( \alpha \), \( u \), and \( v \), respectively, such that \({\theta}_{B}=\{{\theta}_{B, \alpha},  {\theta}_{B, u}, {\theta}_{B, v}\} \). The networks are trained by minimizing the following loss functions:
\begin{equation}\label{eq:loss_branch}
\begin{aligned}
\mathcal{L}_{B, \alpha}({\theta}_{B, \alpha}) &= \left\| {B}_{\alpha}^T({\theta}_{B, \alpha}) - {R}^* {A}_{\alpha}^* \right\|_{L^2}, \\
\mathcal{L}_{B, u}({\theta}_{B, u}) &= \left\| {B}_{u}^T({\theta}_{B, u}) - {R}^* {A}_{u}^* \right\|_{L^2}, \\
\mathcal{L}_{B, v}({\theta}_{B, v}) &= \left\| {B}_{v}^T({\theta}_{B, v}) - {R}^* {A}_{v}^* \right\|_{L^2},
\end{aligned}
\end{equation}
where 
\({B}_{\alpha}({\theta}_{B, \alpha})\), \({B}_u({\theta}_{B, u})\), and \({B}_v({\theta}_{B, v}) \in \mathbb{R}^{m \times p}\)
represent the output of the branch networks corresponding to \( \alpha \), \( u \), and \( v \), respectively. 

Upon training both the trunk and branch networks, the final prediction of the DeepONet model is constructed by combining the outputs of the trained trunk and branch networks following \eqref{eq:deeponet_outputs7}. This sequential training approach enables the model to refine its predictions iteratively: first by learning a structured base through the trunk network and subsequently enhancing the output using the branch network.


\subsection{Physics-informed two-step DeepONet}\label{sec:Physics informed Two step Deepont}

Incorporating physics into the DeepONet framework enhances its ability to model complex systems and makes the network prediction robust, accurate, and adhering to the governing physical laws. In a physics-informed two-step DeepONet, physics is enforced by modifying the trunk network's loss function as follows:
\begin{equation}\label{eq:total_loss}
\begin{aligned}
\mathcal{L}_T({\theta}_T, {A}_{\alpha}, {A}_{u}, {A}_{v}) &= 
\lambda_{\text{data}} \, \texttt{loss}_{\text{data}} + \lambda_{\text{physics}} \, \texttt{loss}_{\text{physics}} \,, \\
&= \lambda_{\text{data}} \, \mathcal{L}_{T_{\text{data}}}({\theta}_T, {A}_{\alpha}, {A}_{u}, {A}_{v}) + 
\lambda_{\text{physics}} \, \mathcal{E}(\bm{u}_{\theta}, \alpha_{\theta}).
\end{aligned}
\end{equation}
where \(\bm{u}_{\theta}\) is the displacement field with components \( u_{\theta} \) and \( v_{\theta} \), and \(\lambda_{\text{data}}\) and \(\lambda_{\text{physics}}\) are weighting factors that balance the importance of physical constraints and the data-driven loss. 
Subsequently, the branch network is trained using the trunk network’s output as described in Section~\ref{Two-step}.

\subsection{Operator network design and training}\label{sec:Operator design}


In this work,  we employ the robust two-step DeepONet training approach~\cite{lee2024training}, sketched in Figure~\ref{fig:Two_step_uvalpha}, using MLPs for both the trunk and branch networks. In the first approach of data-driven two-step DeepONet training, the trunk network loss function includes only the data loss, with no physics-informed component. In the second approach of physics-informed two-step DeepONet training, the trunk network loss function incorporates physics-based loss terms in addition to the data loss. In training the DeepOKAN, we do not employ the two-step training and train the branch and trunk networks concurrently, with the trunk network implemented as a KAN and the branch as an MLP. In all approaches, the networks are trained using the data generated from FE simulations. See~\ref{sec:Tdg} for the details of the FE computations.

In the physics-informed training, the computation of \(\mathcal{E}\) and its minimization poses challenges due to the nonconvexity and complexity of the energy functional, requiring careful optimization. To efficiently compute and minimize the energy functional, we utilize the following techniques:
\begin{itemize}
    \item Instead of employing automatic differentiation to compute gradients, we mesh the domain and compute the gradients of the fields as well as perform the integration in \eqref{eq:2} following the computational approach based on the finite element (FE) method described in~\ref{sec:Energy computation}. This approach provides a computationally efficient and geometrically intuitive way to capture spatial variations in field values~\cite{manav2024phase}.  
    Note that the computation of energy and its minimization requires the evaluation of the solution fields across the entire spatial domain simultaneously.

    \item In training, we employed an adaptive weighting method to balance between $\texttt{loss}_{\alpha}$, $\texttt{loss}_{u}$, $\texttt{loss}_{v}$, and $\texttt{loss}_{\text{physics}}$. Initially, $\texttt{loss}_{\text{physics}}$ was much larger than $\texttt{loss}_{\text{data}}$, necessitating the introduction of adaptive weights to effectively balance the two loss functions. For details on the adaptive weighting scheme, please refer to~\ref{sec:Adaptive weighing} and see~\cite{kendall2018multi,li2022revisiting,chen2024self}.
\end{itemize}

\section{Prediction of fracture with data-driven two-step DeepONets}\label{sec:Data-Driven}

We demonstrate the performance of the proposed networks for two sets of problems: (a) a one-dimensional homogeneous bar with prescribed displacement at the boundaries, and (b) two-dimensional single-edge notched specimens with different boundary conditions and modeling assumptions. For the bar problem, the prescribed boundary displacement is the input to the branch network. For the problems with the notched specimen, the length of the initial crack-like notch, represented by an initial phase field, is the input to the branch network. Leveraging the monotonically increasing discrete loading procedure, we train separate networks to represent the solution fields at different loading levels, i.e. for each prescribed displacement \( U_t,~t\in\{0,1,2,...\}\) (with \(U_0=0\)), we train a network that approximates the operator \(\mathcal{G}_t:\mathcal{A}_0\mapsto \bar{G}_t\) where \(\mathcal{A}_0\) is the set of phase fields corresponding to the notches of different sizes and \( \bar{G}_t\) is the set of the solution fields, that is the displacement and phase fields, at time $t$. Since the input to \(\mathcal{G}_t\) is \(\alpha_0\in \mathcal{A}_0 \) and not the solution fields in the previous loading step, unlike in previous work~\cite{GOSWAMI2022114587}, our approach does not suffer from the problem of error accumulation.

In the data-driven framework, the trunk network is trained using only the data loss in~\eqref{eq:loss_trunk} corresponding to the displacement and phase fields. 
We start by studying crack nucleation in a one-dimensional homogeneous bar (\textbf{Case 1}). Subsequently, we study crack propagation in single-edge notched (SEN) specimens subjected to tensile loading (\textbf{Case 2}), and crack propagation with kinking in the same specimens under shear loading (\textbf{Case 3}). We also study crack branching in the same specimens, again under shear loading (\textbf{Case 4}). For the first two cases,  the one-dimensional bar and the SEN specimens with tensile loading, FE data is generated using the AT2 damage model, which leads to an immediate onset of damage at the beginning of loading. For the next two cases, the SEN specimens with shear loading leading to crack kinking and branching, we generate FE data using the AT1 damage model, which leads to a linear-elastic behavior with no damage evolution in the initial stage of loading and the initiation of damage evolution only after a load threshold is reached. This is a comparatively challenging case for DeepONet to learn. These two examples demonstrate the versatility of the framework in handling both sharp and gradual damage evolution behaviors, thus encompassing a broader range of scenarios encountered in fracture modeling.
For all examples, unless otherwise specified, the networks are trained using the Adam optimizer~\cite{kingma2014adam} with a learning rate of \( 10^{-4} \), a mean squared error (MSE) loss function, and the \texttt{tanh} activation function is used throughout the architecture.

\begin{table}[ht]
\centering
\begin{tabular}{|l|l|l|}
\hline
     & \textbf{Name}    & N\textsubscript{s}  
\\ \hline
\textbf{Case 1}  & One-dimensional homogeneous bar          & 101  
\\ \hline
\textbf{Case 2}  & SEN under tensile loading    & 33749
\\ \hline
\textbf{Case 3}  & SEN under shear loading      & 149383    
\\ \hline
\textbf{Case 4}  & SEN under shear loading (branching)                   & 252996     
\\ \hline
\end{tabular}
\caption{Summary of the examples discussed in this study. In this table, N\textsubscript{s} represents the number of sensor points given as input to the trunk network.}
\label{table_examples}
\end{table}

\subsection{One-dimensional homogeneous bar (\textbf{Case 1})}

\begin{figure}[!tbh]
    \centering
    \includegraphics[width=0.3\textwidth]{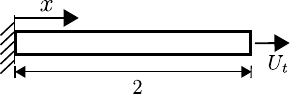}
    \caption{\textbf{Case 1}: One-dimensional homogeneous bar: geometry and boundary conditions.}
    \label{fig:1Dbar}
\end{figure}

In this section, we study the one-dimensional homogeneous bar shown in Figure~\ref{fig:1Dbar}. Here, we set \(E=1.0\), \(G_c=0.01\), and \(l = 0.1\). The data set comprises the phase and displacement fields at 101 equispaced nodes in the discretization of the domain for 50 equispaced applied displacements such that \( U_t \in \{0.01, 0.02, \cdots, 0.5\}\). The data set is divided into training and test sets, with the data for the first 45 values of \( U_t \) used for training and the data for the last 5 values of \( U_t \) reserved for testing. A two-step DeepONet architecture is employed, consisting of a branch network, which processes the applied displacement values, and a trunk network, which operates on the coordinates of the nodes. The trunk network has three hidden layers, each containing 101 neurons, while the branch network consists of five hidden layers, each with 101 neurons. The dimension of the outputs of both the networks is 45.
The network is trained to predict the phase field \( \alpha \) and the displacement field \( u \) with a learning rate of \(10^{-5} \).

\begin{figure}[H]
    \centering
    \includegraphics[width=0.79\textwidth]{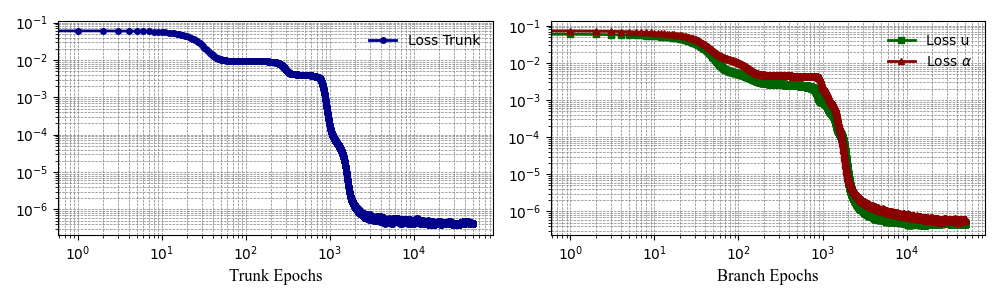}
    \caption{\textbf{Case 1}: The evolution of the total loss (losses associated with both \( u \) and \( \alpha \)) for the trunk network during training eventually reaching \( 10^{-6} \). Similarly, the loss function for the branch network, responsible for predicting both \( u \) and \( \alpha \), also converges to \( 10^{-6} \).}   
    \label{fig:loss_branchnet1}
\end{figure}
The plots in Figures~\ref{fig:loss_branchnet1} show the evolution of the losses for the trunk and branch nets during training. The loss in the trunk net is a combination of the losses associated with \( u \) and \( \alpha \). Losses associated with \( u \) and \( \alpha \) in the trunk network training are closely aligned. The convergence of the losses for both the networks to very small values reveals the success of the training approach. 

\begin{figure}[!tbh]
\centering
\includegraphics[width=0.995\textwidth]{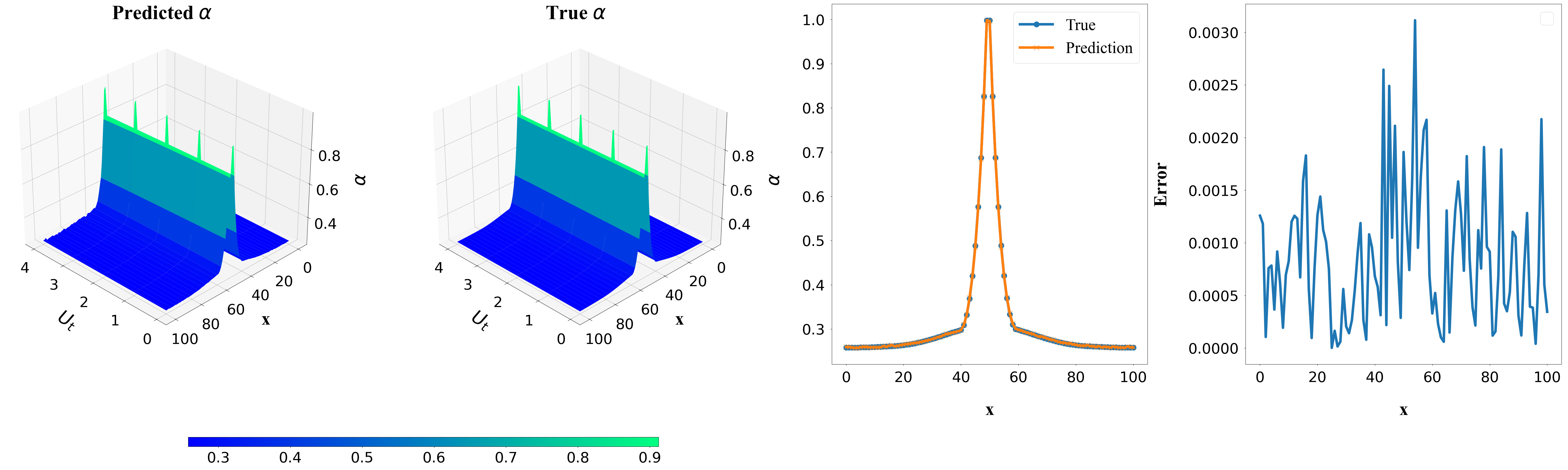}
\caption{\textbf{Case 1}: Comparison of the true \( \alpha \) and the predicted \( \alpha \) by the data-driven two-step DeepONet. The figure on the right shows the absolute error in prediction.}
\label{fig:alpha_predalpha1}
\end{figure}

\begin{figure}[!tbh]
    \centering
    \includegraphics[width=0.995\textwidth]{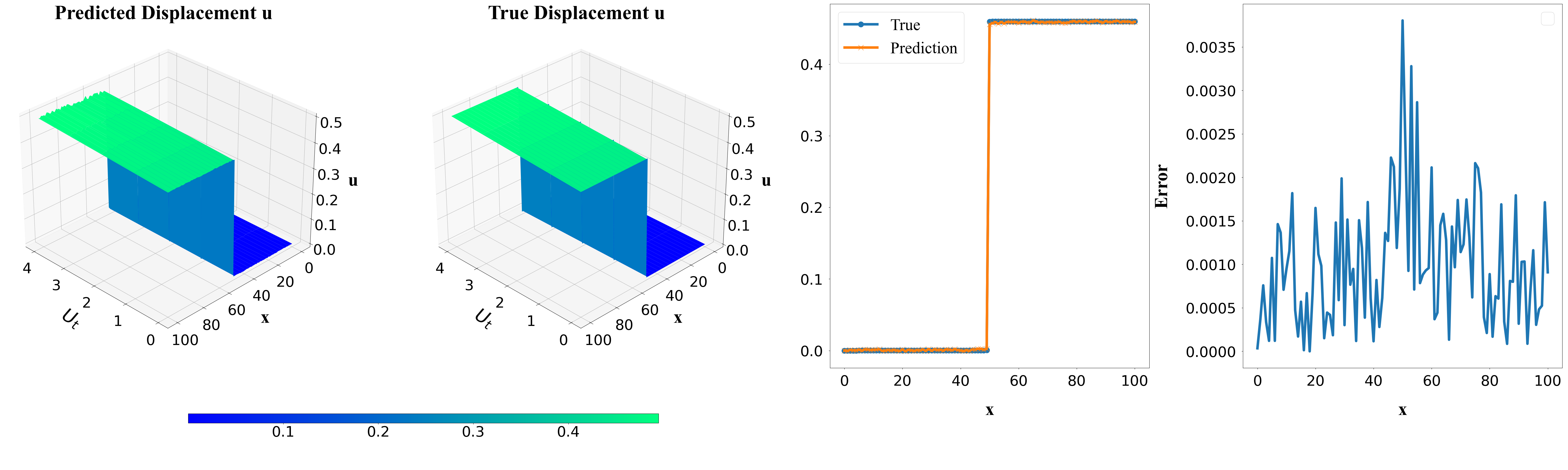}
    \caption{\textbf{Case 1}: Comparison of the true displacement and the displacement predicted by the data-driven two-step DeepONet, along with the absolute error in prediction.}
    \label{fig:u_prediction1}
\end{figure}

The plots in Figures~\ref{fig:alpha_predalpha1} and~\ref{fig:u_prediction1} illustrate the evolution of the phase field \( \alpha \) and the displacement field \( u \) versus the applied displacement values \( U_t \). The predictions are displayed for five incremental steps of \( U_t \), offering a detailed comparison between the predicted values and the ground truth. The performance of the neural networks is quantitatively evaluated by analyzing the absolute errors, which are reported to be less than 0.0025 for \( \alpha \) and less than 0.0035 for \( u \). These results emphasize the high accuracy of the model in capturing the intricate behaviors of both the phase and displacement fields, including very sharp gradients.

\subsection{\textbf{Case 2}: Crack propagation in an SEN
specimen under tensile loading}
\begin{figure}[H]
    \centering
    \includegraphics[width=0.3\textwidth]{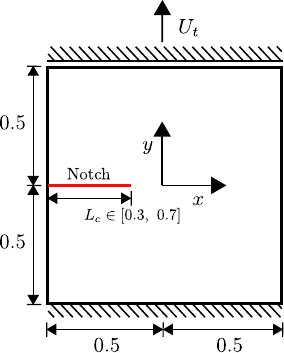}
    \caption{\textbf{Case 2}: SEN specimen under tensile loading: geometry and boundary conditions.}
    \label{fig:SENtensile}
\end{figure}
In this example, we demonstrate the ability of DeepONet to predict crack propagation by training the network on FE data for SEN specimens with varying initial notch sizes subjected to tensile loading~(Figure~\ref{fig:SENtensile}). For this problem, we set \(E=1.0\), \(\nu=0.3\), \(G_c=1.0\), and \(l = 0.01\). In this case, the branch network processes the phase field associated with the initial notches, while the trunk network receives the spatial coordinates of 33749 points as input. The trunk network is composed of 7 hidden layers, each containing 1001 neurons, while the branch network comprises 9 hidden layers and 1001 neurons. The final layer of each network is configured with 45 neurons. The dataset consists of the displacement and phase fields for 50 specimens, each with a distinct initial notch size, and each subjected to 40 equispaced applied displacements such that \( U_t \in \{0.005, 0.01, \cdots, 0.2\}\). The data for 45 randomly selected specimens are used for training, while the data for the remaining 5 specimens are reserved for testing.

\begin{figure}[H]
    \centering
    \includegraphics[width=0.99\textwidth]{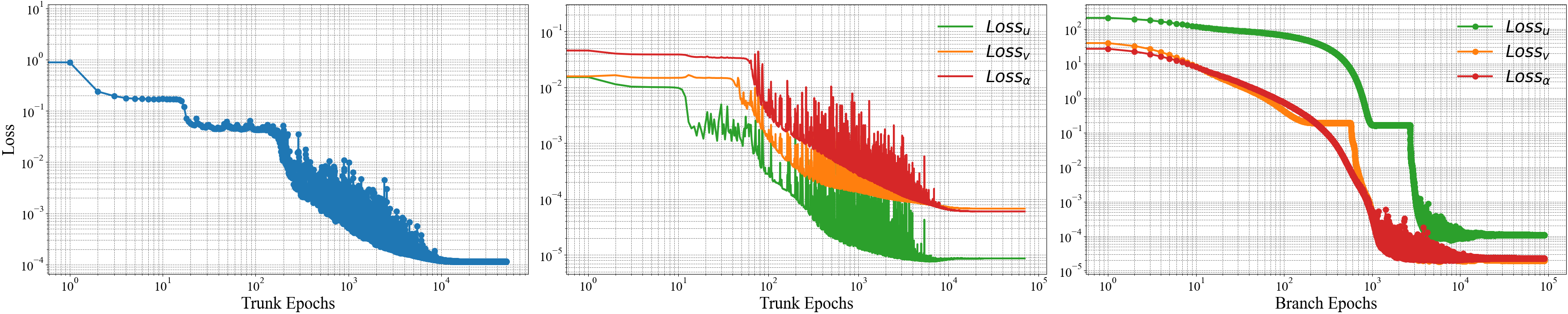}
\caption{\textbf{Case 2}: The  evolution of the loss during the training of the trunk and branch networks for the problem of the SEN specimens under tensile loading conditions. The middle figure shows the adjusted losses after applying \( \lambda_{\alpha} \), \( \lambda_{u} \), and \( \lambda_{v} \), which are modified to balance the contributions of each loss. Since \( \texttt{loss}_{u} \) in the trunk network is relatively small, the corresponding weight \( \lambda_{u} \) controls its influence. Conversely, because \( \texttt{loss}_{\alpha} \) is larger than the other losses, \( \lambda_{\alpha} \) is adjusted to ensure a balanced contribution across all components.
}
    \label{fig:loss_weights_tensil}
\end{figure}

Figure~\ref{fig:loss_weights_tensil} shows the training losses for both the trunk and the branch networks. During trunk network training, the loss function shows a consistent decrease, eventually reaching a value of the order of \(10^{-4}\), 
suggesting that training has reached convergence.
Notably, in the early stages of training, a pronounced imbalance was observed among the three components of loss (\( \texttt{loss}_{\alpha} \), \( \texttt{loss}_{v} \), and \( \texttt{loss}_{u} \)), with \( \texttt{loss}_{\alpha} \) significantly larger than \( \texttt{loss}_{u} \) and \( \texttt{loss}_{v} \), the latter being particularly small. To address this imbalance and to ensure that each loss component contributes meaningfully to the optimization process, we implemented an adaptive weight adjustment mechanism.
This mechanism dynamically adjusts the weights of each loss component (\( \lambda_{u} \), \( \lambda_{v} \), and \( \lambda_{\alpha} \)) by normalizing their contributions relative to the total loss and applying adjustments to maintain numerical stability. By facilitating learning from all loss components, it enhances the overall performance and robustness of the method. The middle panel of Figure~\ref{fig:loss_weights_tensil} demonstrates the balanced losses achieved by applying dynamic weighing. The figure highlights that \( \texttt{loss}_{\alpha} \) is initially larger than the other losses but progressively decreases during training, eventually aligning more closely with \( \texttt{loss}_{u} \) and \( \texttt{loss}_{v} \). Additionally, \( \texttt{loss}_{u} \) remains relatively small compared to the other loss components throughout the training process.

\begin{figure}[H]
    \centering
    \begin{subfigure}[b]{0.510\textwidth}
        \centering
        \includegraphics[width=\textwidth]{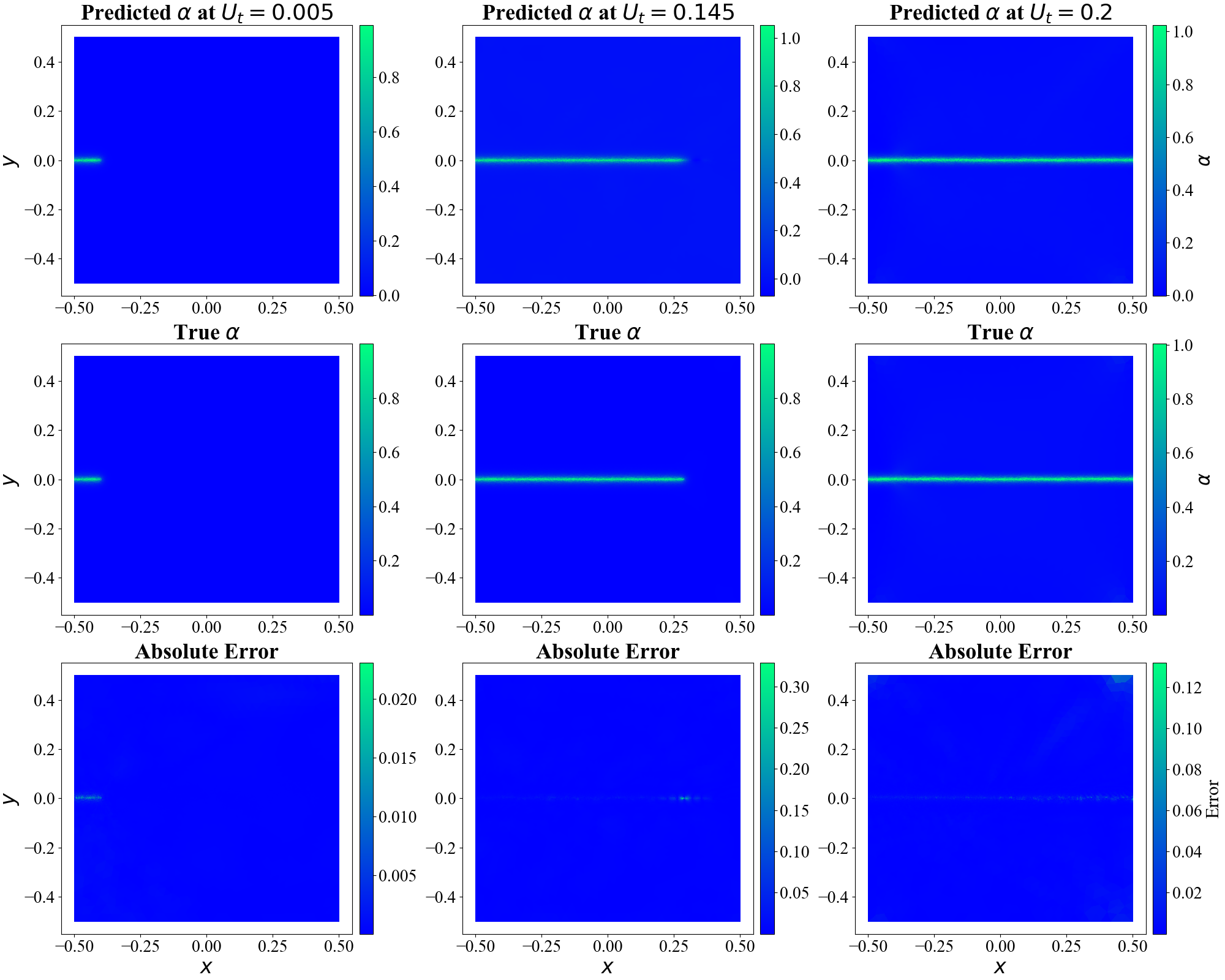}
        \caption{Predicted damage \(\alpha\)}
    \end{subfigure}
    \hfill
    \begin{subfigure}[b]{0.495\textwidth}
        \centering
        \includegraphics[width=\textwidth]{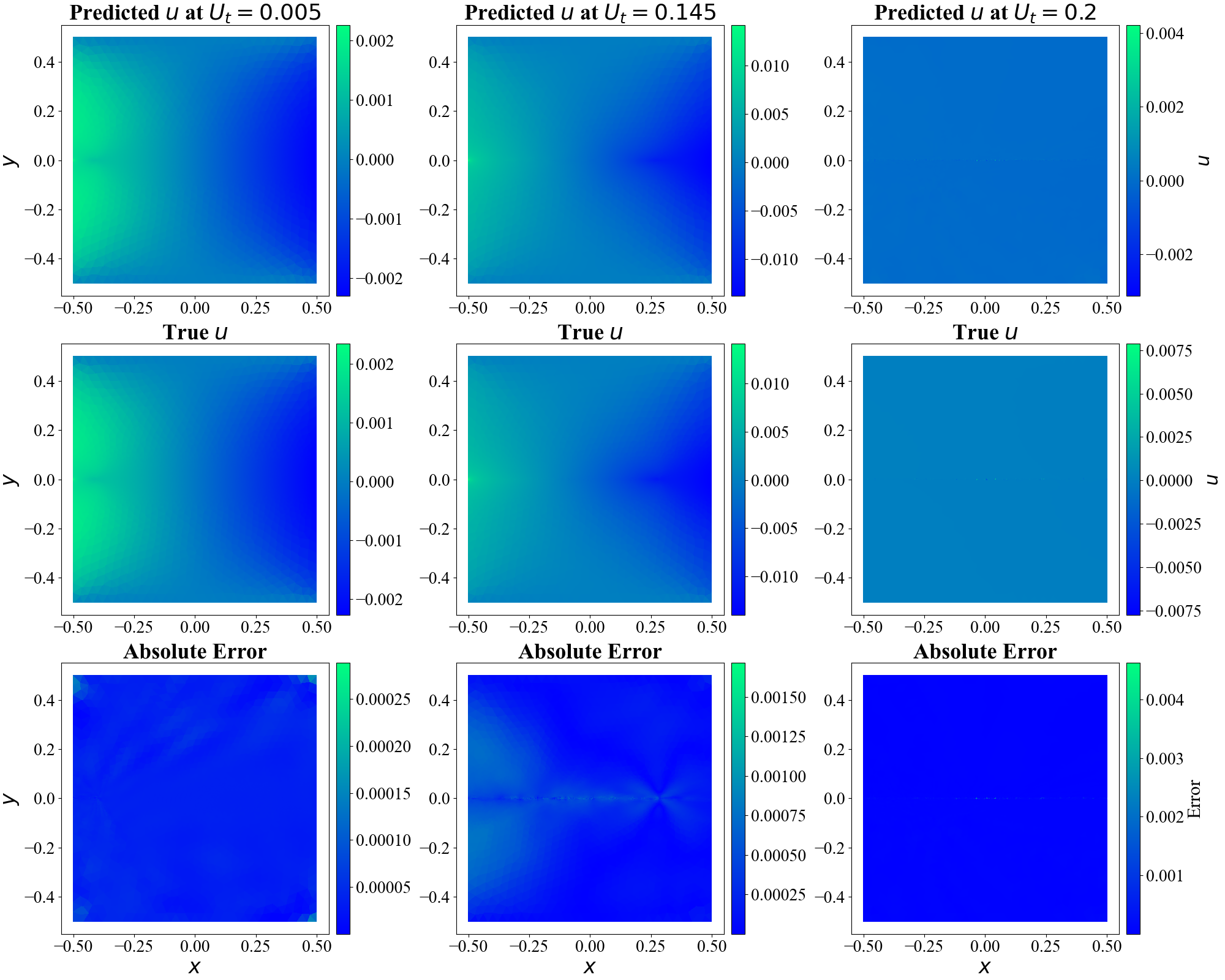}
        \caption{Predicted displacement \(u\)}
    \end{subfigure}
    \hfill
    \begin{subfigure}[b]{0.495\textwidth}
        \centering
        \includegraphics[width=\textwidth]{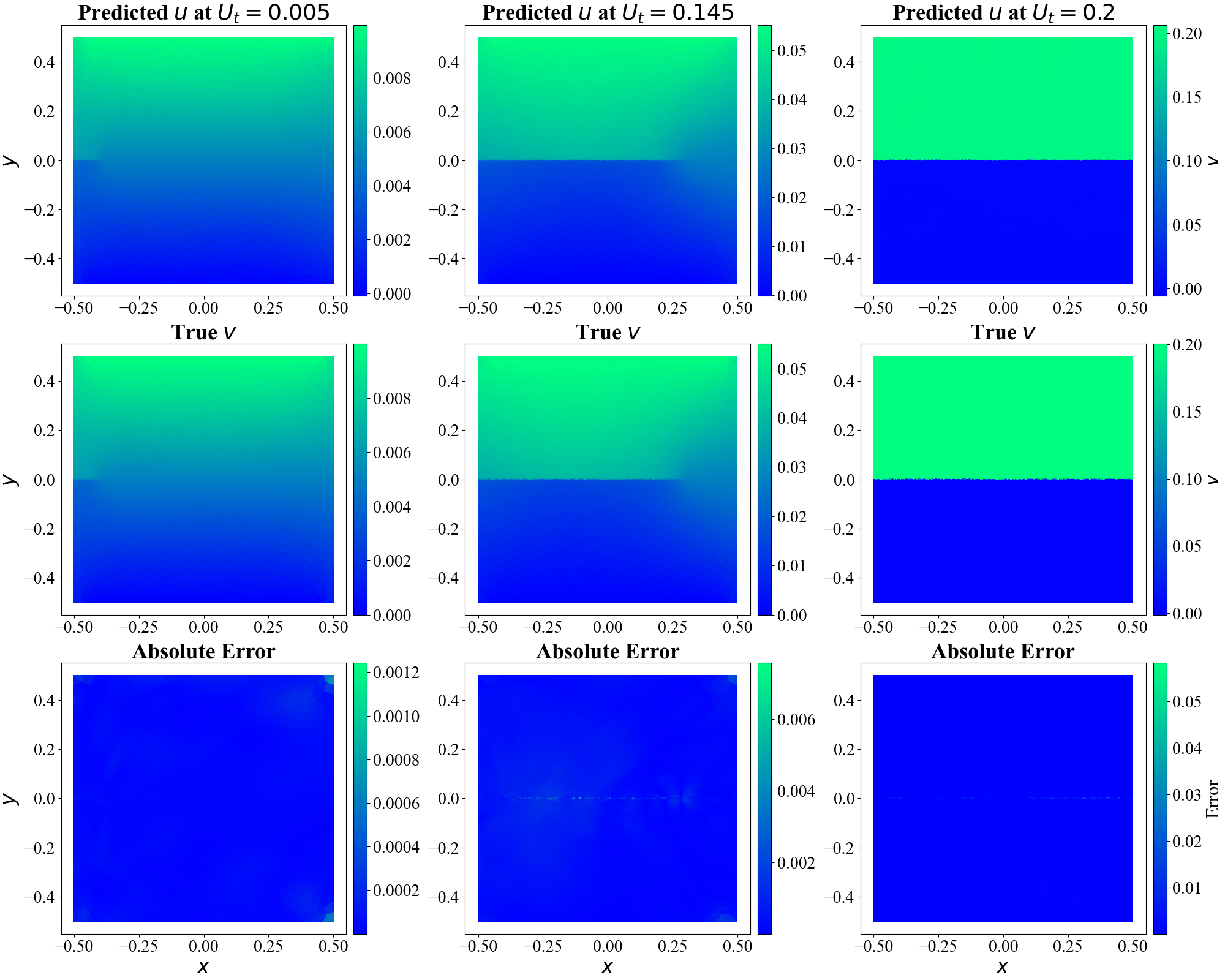}
        \caption{Predicted displacement \(v\)}
    \end{subfigure}
     \caption{\textbf{Case 2}: Comparison of the true phase and displacement fields with the fields predicted by the data-driven two-step DeepONet for three applied displacement values: \(U_t = 0.005\), \(U_t = 0.145\), and \(U_t = 0.2\). In each plot, the first row shows the predictions, the second row shows the true values, and the bottom row illustrates the absolute error.}
    \label{fig:u_prediction2_tensil}
\end{figure}

Figures~\ref{fig:u_prediction2_tensil} compares the predicted phase and displacement fields  at the applied displacement values of
\(U_t = 0.05\), \(U_t = 0.145\), and \(U_t = 0.2\) for one of the test cases. 
For \(U_t = 0.05\) and \(U_t = 0.2\), the phase and displacement fields are predicted accurately with the error localized to only a few points near the crack. For \(U_t = 0.145\), the phase field in front of the crack tip decays smoothly unlike in the FE data, pointing to the challenges in learning these complex fields. Overall, the networks predict the fields with high accuracy, highlighting the generalization capability of the model, achieved using only 45 training samples.

\subsection{\textbf{Case 3}: Crack propagation in a SEN specimen under shear loading}
\label{case3}
\begin{figure}[H]
    \centering
    \includegraphics[width=0.3\textwidth]{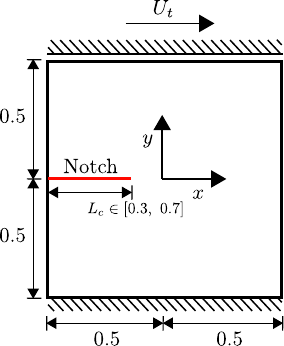}
    \caption{\textbf{Case 3}: SEN specimen under shear loading: geometry and boundary conditions.}
    \label{fig:SENshear}
\end{figure}
In this example, the ability of the network to predict crack kinking is demonstrated using a network trained on FE data for SEN specimens with varying initial notch lengths subjected to shear loading~(Figure~\ref{fig:SENshear}). For this problem also, we set \(E=1.0\), \(\nu=0.3\), \(G_c=1.0\), and \(l = 0.01\). Here, the branch network processes the phase field associated with the initial notches and the trunk network processes 149383 sensor points. The structure of the trunk network and branch network  follows the same architecture as in \textbf{Case 2}. The dataset consists of the displacement and phase fields for 50 specimens, each with a distinct initial notch size, and each subjected to 50 equispaced applied displacements such that \( U_t \in \{0.01, 0.02, \cdots, 0.5\}\). 
Data from 45 randomly selected initial notch sizes are utilized for training, while data from the remaining 5 initial notch sizes are allocated for testing. The phase and displacement fields are predicted for all 50 applied displacements.

\begin{figure}[H]
    \centering
    \begin{subfigure}[b]{0.510\textwidth}
        \centering
        \includegraphics[width=\textwidth]{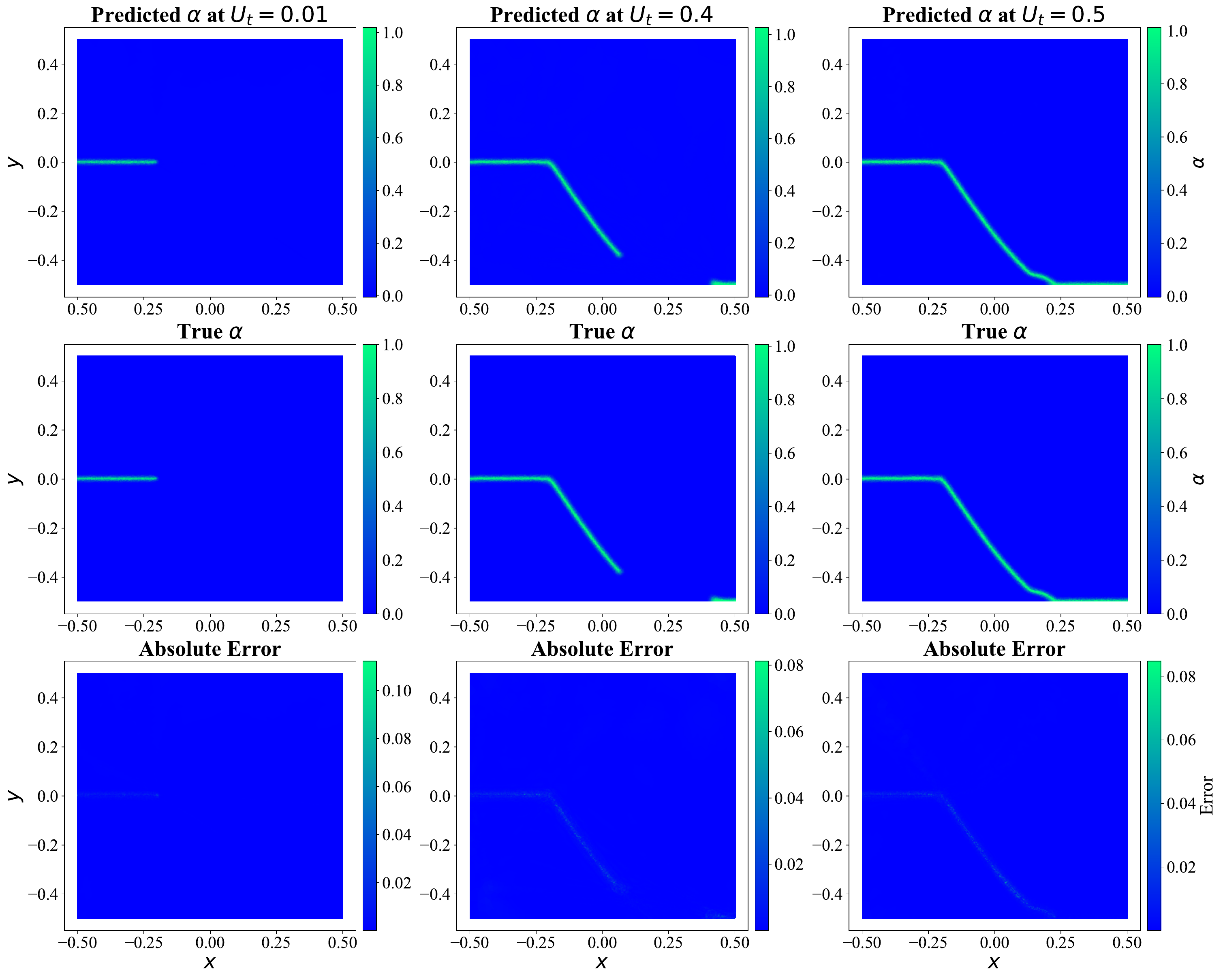}
        \caption{Predicted damage \(\alpha\)}
    \end{subfigure}
    \hfill
    \begin{subfigure}[b]{0.495\textwidth}
        \centering
        \includegraphics[width=\textwidth]{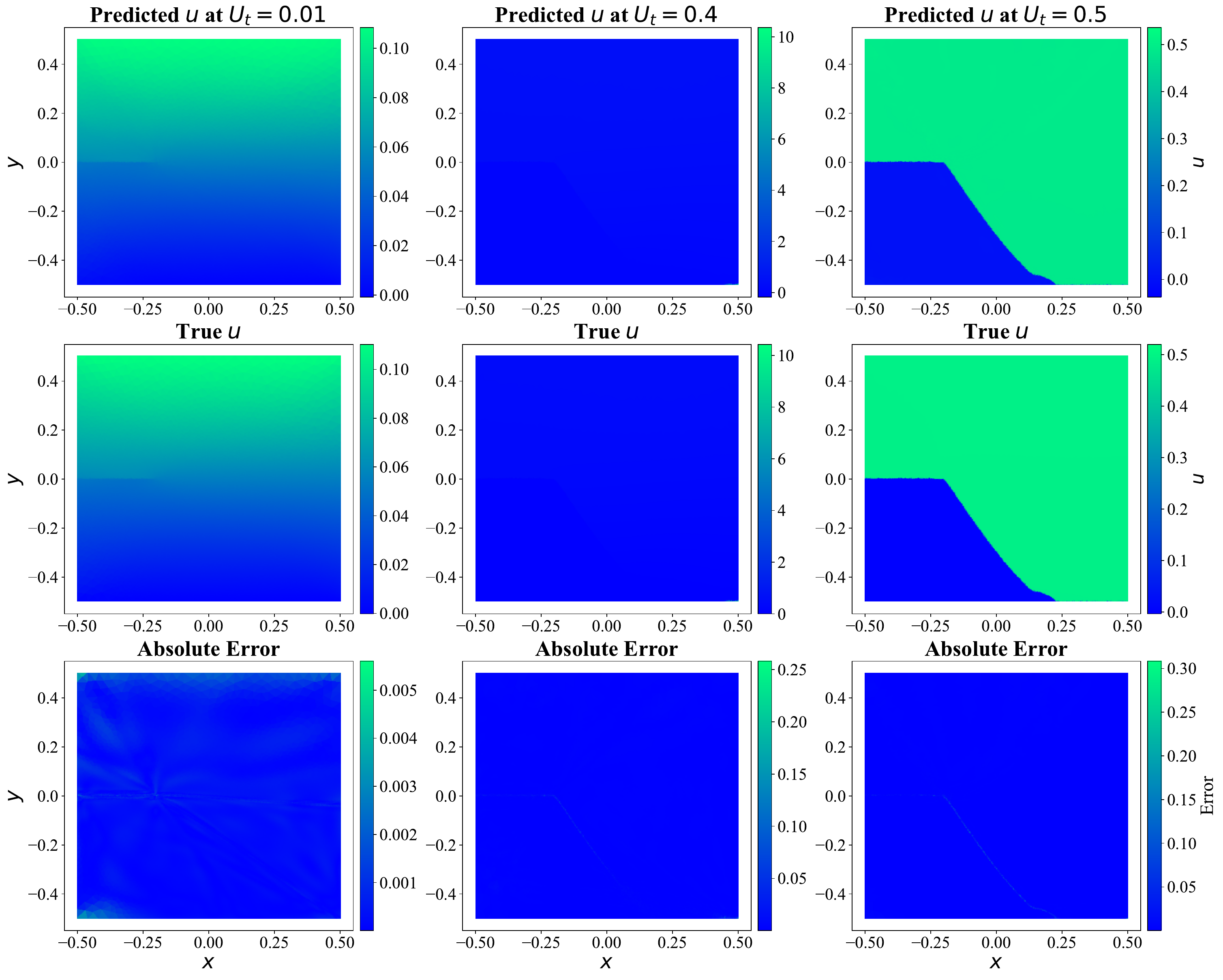}
        \caption{Predicted displacement \(u\)}
    \end{subfigure}
    \hfill
    \begin{subfigure}[b]{0.495\textwidth}
        \centering
        \includegraphics[width=\textwidth]{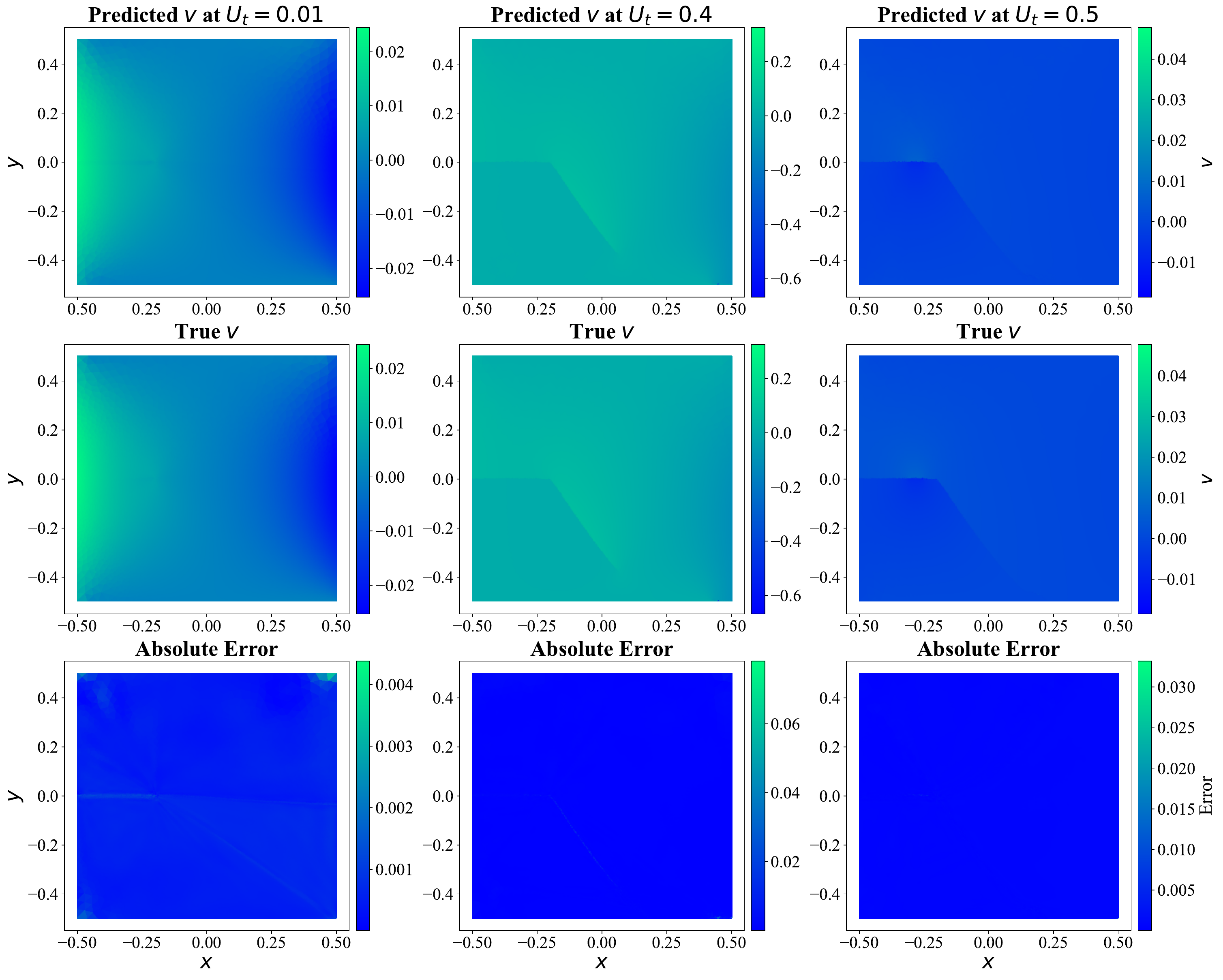}
        \caption{Predicted displacement \(v\)}
    \end{subfigure}       
    \caption{\textbf{Case 3}: 
    Comparison of the true phase and displacement fields with the fields predicted by the data-driven two-step DeepONet for three applied displacement values: \(U_t = 0.01\), \(U_t = 0.4\), and \(U_t = 0.5\). 
    In each plot, the first row shows the predictions, the second row shows the true values, and the bottom row illustrates the absolute error.
    }
    \label{fig:data_driven_SEN}
\end{figure}

Figure~\ref{fig:data_driven_SEN} displays the predicted phase and displacement fields at the applied displacement values of \(U_t = 0.01\), \(U_t = 0.4\), and \(U_t = 0.5\) for one of the test specimens. The networks accurately predict the propagation of a crack with kinking. Remarkably, the network also predicts the nucleation of a crack at the bottom right corner of the specimen, as observed from the plot for \(U_t = 0.4\)). Furthermore, unlike in the case with tensile loading, the phase field does not exhibit an unrealistic smooth decay at the crack tip. Small errors in the predicted fields suggest that the network accurately approximates the operator.

\subsection{\textbf{Case 4}: SEN under shear loading (branching)}
This example demonstrates the ability of the network to predict crack branching. For FE simulation of crack branching under shear loading of the same SEN described in Section~\ref{case3}, we purposefully do not apply the decomposition of the strain energy density in \eqref{eq:2} such that \( \Psi^+ = \Psi\) and \( \Psi^- = 0\). Although this is an artificial example (in the sense that the induced branching is unphysical), it demonstrates the capability of the approach to learn and predict crack branching. The dataset consists of displacement and phase fields for 50 specimens, each with a distinct initial notch size, and each subjected to 50 uniformly spaced applied displacements, \( U_t \in \{0.01, 0.02, \cdots, 0.5\}\).
In this case, similar to \textbf{Case 2} and \textbf{Case 3}, the branch network processes the phase field data corresponding to 45 randomly selected initial notches for training and the remaining 5 samples for testing, while the trunk network processes coordinate points.
The architecture of the network is very similar to that in \textbf{Case 2} and \textbf{Case 3} utilizing the same number of layers and neurons for both the trunk and branch networks.  However, the last layer of each network now consists of 101 neurons.  

\begin{figure}[!tbh]
    \centering
    \begin{subfigure}[b]{0.510\textwidth}
        \centering
        \includegraphics[width=\textwidth]{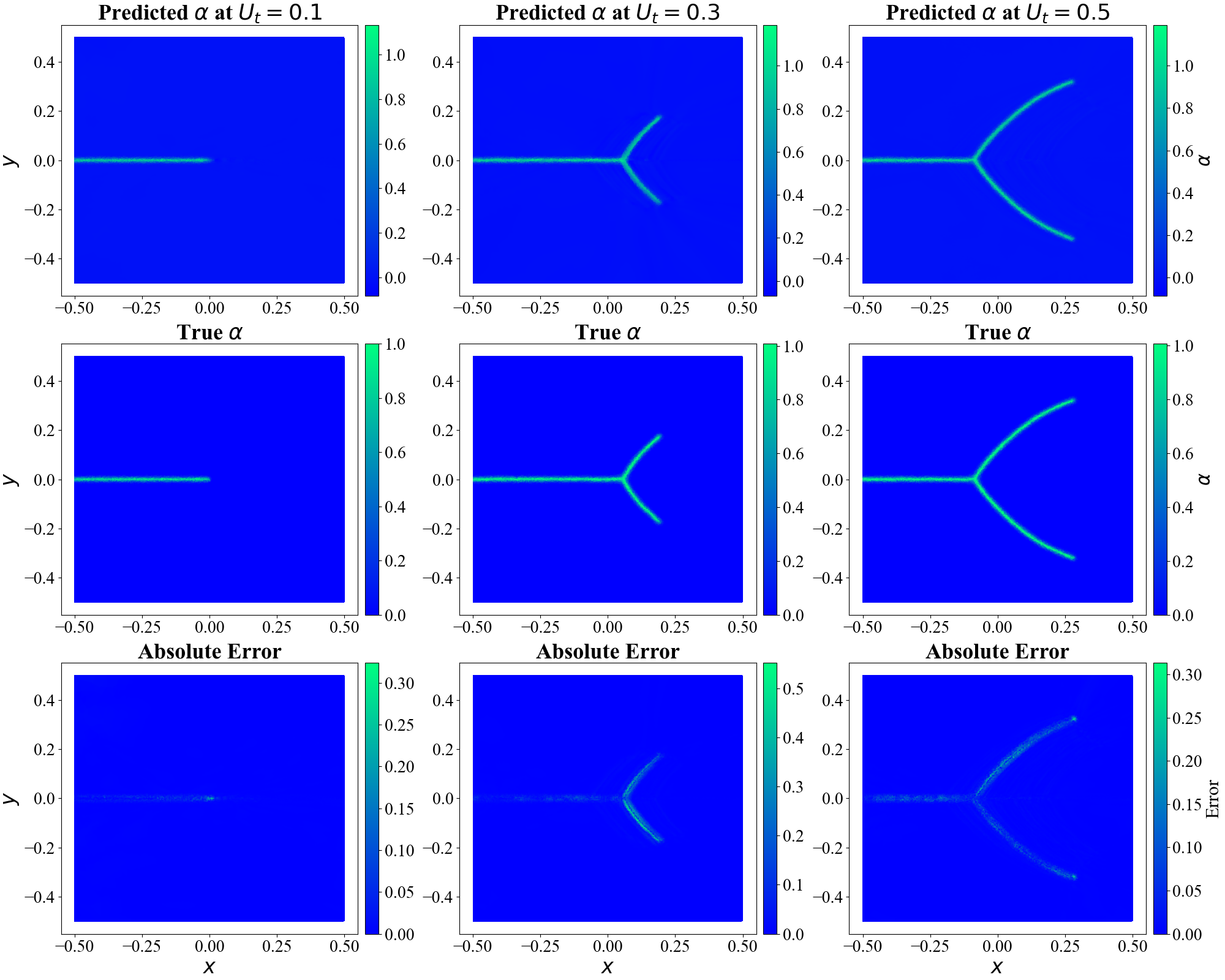}
        \caption{Predicted damage \(\alpha\)}
    \end{subfigure}
    \hfill
    \begin{subfigure}[b]{0.495\textwidth}
        \centering
        \includegraphics[width=\textwidth]{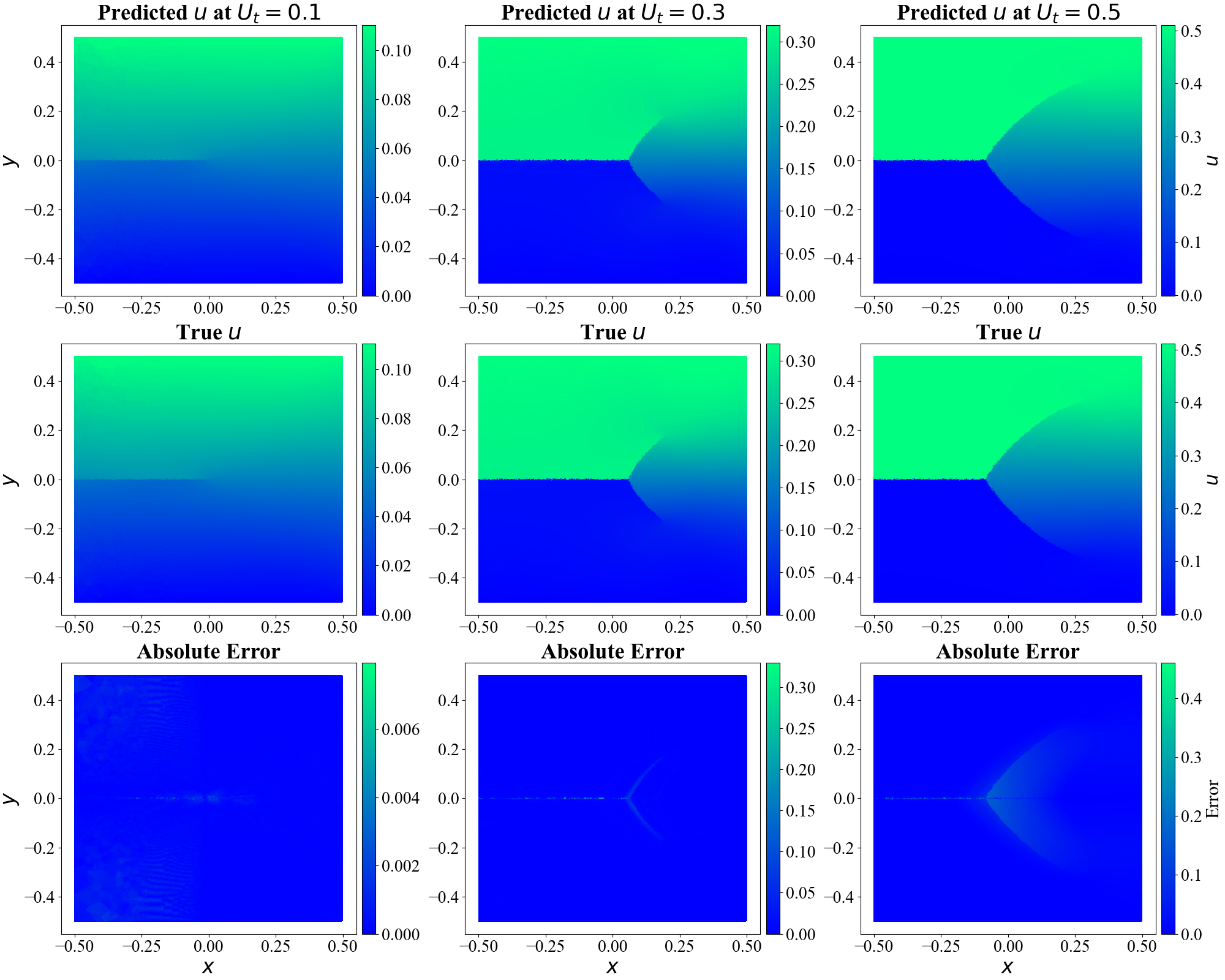}
        \caption{Predicted displacement \(u\)}
    \end{subfigure}
    \hfill
    \begin{subfigure}[b]{0.495\textwidth}
        \centering
        \includegraphics[width=\textwidth]{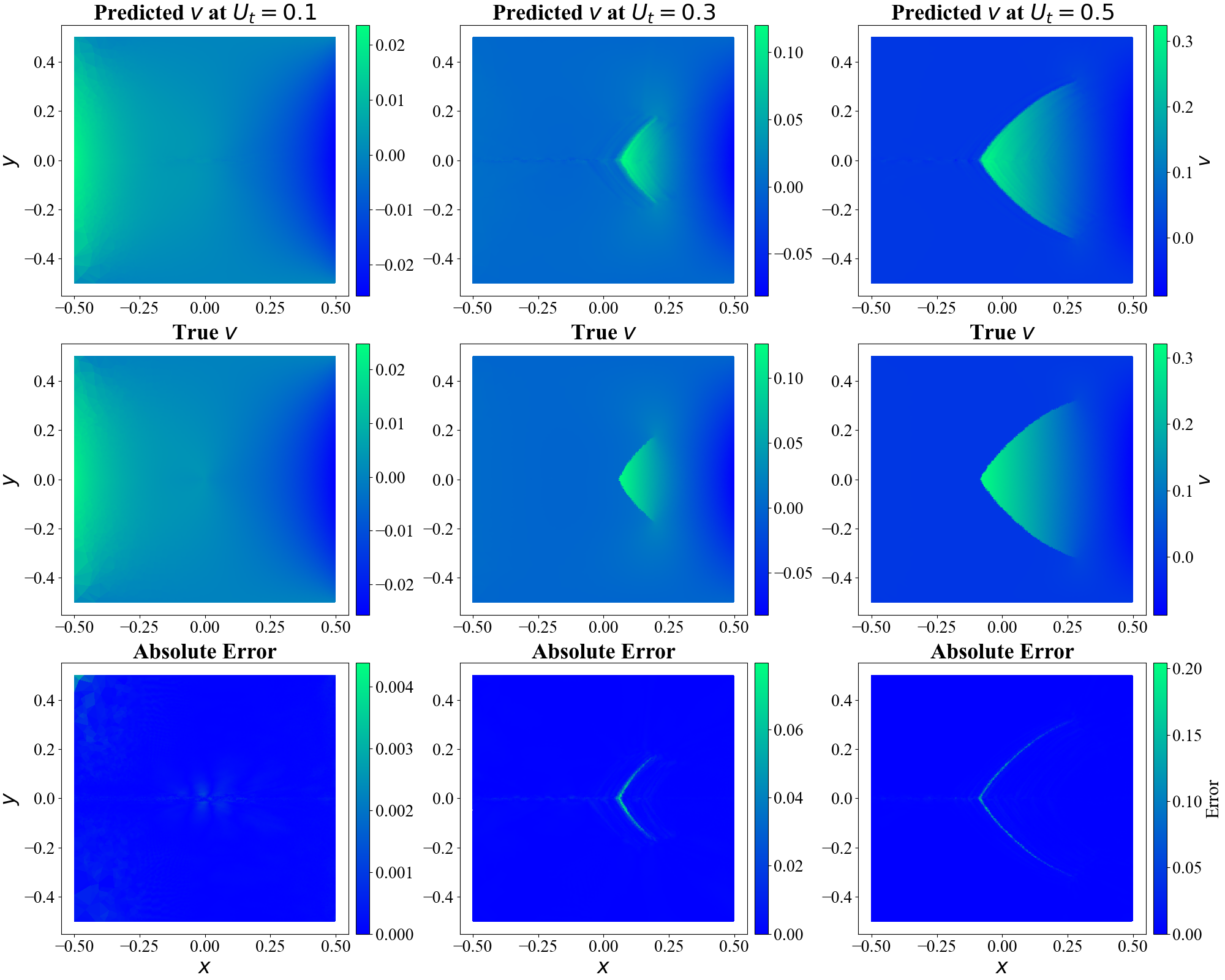}
        \caption{Predicted displacement \(v\)}
    \end{subfigure}
\caption{\textbf{Case 4}: Comparison of the predicted (top row) and true (middle row) fields for (a) damage \(\alpha\), (b) displacement \(u\), and (c) displacement \(v\) at three applied displacement values: \(U_t = 0.1\), \(U_t = 0.3\), and \(U_t = 0.5\), representing crack propagation with branching. 
The bottom row in each plot shows the absolute error between the predicted and true fields.}
    \label{fig:u_branching}
    \end{figure}

The performance of the two-step DeepONet in predicting crack propagation with branching is illustrated in Figure~\ref{fig:u_branching}, which displays predicted \(\alpha\), \(u\), and \(v\). The network predicts the crack path correctly. However, while the fields away from crack are predicted accurately, some error is observed in the vicinity of the crack, pointing to difficulties in learning the sharply varying fields near the crack accurately.

\subsection{Accuracy of predictions}
The mean absolute errors (MAEs) using the 5 specimens reserved for testing for the three two-dimensional cases—tensile (\textbf{Case 2}), shear (\textbf{Case 3}), and shear with branching (\textbf{Case 4})—are presented in Table~\ref{Table:errors_data_driven}. It is evident that the MAEs for the displacement field components \(u\), \(v\), and the phase field \(\alpha\) are generally on the order of \(10^{-3}\) or \(10^{-4}\), suggesting good generalizability of the trained networks.
Table~\ref{table_training_summary_data-driven} reports the number of epochs for both the trunk and branch networks, along with the training time required for predicting each case. In all cases, 45 samples are used for training and 5 samples for testing in the branch network, while the sensor points listed in Table~\ref{table_examples} are used for training the trunk network. Notably, in \textbf{Case 4}, only half of the sensor points from Table~\ref{table_examples} are used for training, whereas in all other cases, the full set of sensor points is utilized.

\begin{table}[h!]
    \centering
    \begin{tabular}{|c|c|c|c|}
        \hline
        \textbf{Case} & \textbf{MAE \(u\)} & \textbf{MAE \(v\)} & \textbf{MAE \(\alpha\)} \\
        \hline
        Tensile (Case 2) & 1.60e-04 & 2.46e-03 & 1.09e-02 \\
        \hline
        Shear (Case 3) & 6.87e-04 & 1.56e-04 & 6.80e-04 \\
        \hline
        Branching (Case 4) & 1.23e-03 & 5.14e-04 & 4.60e-03 \\
        \hline
    \end{tabular}
\caption{\textbf{Data-driven two-step DeepONet}: MAE for  \( u \), \( v \), and \(\alpha\) for \textbf{Case 2}, \textbf{Case 3}, and \textbf{Case 4}.}
    \label{Table:errors_data_driven}
\end{table}

\begin{table}[ht]
\centering
\small
\renewcommand{\arraystretch}{1.2} 
\setlength{\tabcolsep}{2pt} 
\begin{tabular}{|l|c|c|c|c|c|}
\hline
\textbf{Case} &  \textbf{Trunk} & \textbf{Branch} & \textbf{Comp. Time (Trunk)} & \textbf{Comp. Time ($\alpha$)} & \textbf{Comp. Time (u/v)} \\ 
\hline
\textbf{Case 1}   & 500000 epochs & 500000 epochs & 1381 s & 378 s & 386 s \\                   
\hline 
\textbf{Case 2}   & 700000 epochs& 900000 epochs& 22521 s & 1063 s & 1078/1082 s\\                  
\hline
\textbf{Case 3}   & 500000 epochs& 700000 epochs & 95621 s & 2444 s & 2444/2438 s \\ 
\hline
\textbf{Case 4}   & 300000 epochs& 600000 epochs& 16018 s & 1403 s & 1395 s \\                  
\hline
\end{tabular}
\caption{\textbf{Data-driven two-step DeepONet}: Training details for different data-driven cases using two-step DeepONet architectures. The number of sensor points used in the trunk network for each case is listed in Table 1. In all cases, 45 samples are used for training and 5 samples for testing. The results reported were obtained using an NVIDIA H100 GPU.}
\label{table_training_summary_data-driven}
\end{table}

\subsection{Basis functions}
We also investigate the basis functions obtained during the training of the trunk network. 
As detailed in Section~\ref{Two-step}, we employ the QR decomposition to derive the orthonormal basis functions for the trunk network during training. However, visualizing these basis functions directly from the QR decomposition poses practical challenges. To address this, we utilize SVD as an alternative approach. By applying SVD to the output of the trunk network, we extract and plot the basis functions, facilitating better interpretability and visualization.

\begin{figure}[H]
    \centering
    \includegraphics[width=0.99\textwidth]{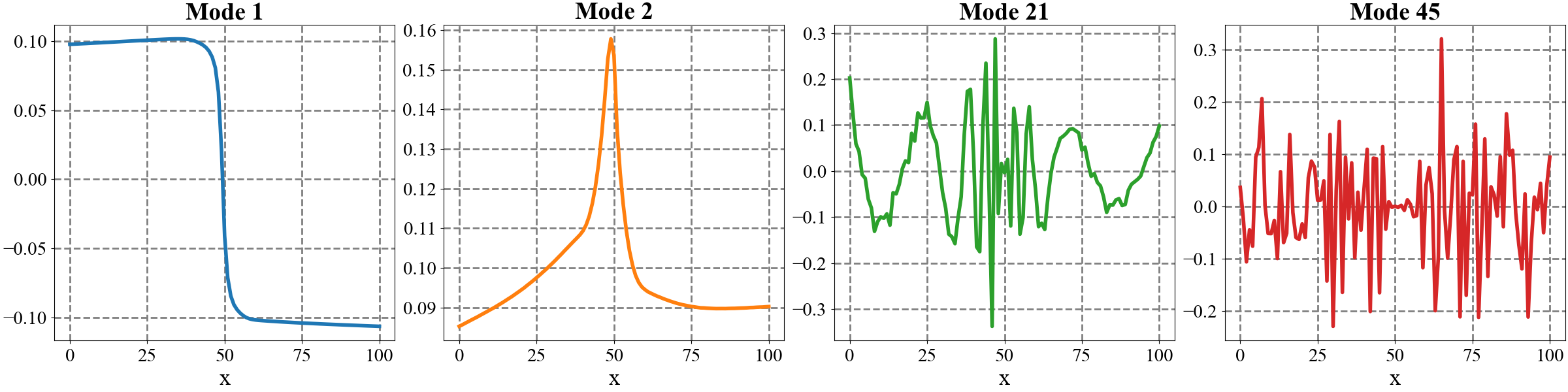}
    \caption{\textbf{Case 1}: Basis functions generated by the trunk network for modes 1, 2, 21, and 45.}
    \label{fig:basis_functions1D}
\end{figure}

\begin{figure}[!tbh]
    \centering
    \begin{subfigure}[b]{0.510\textwidth}
        \centering
        \includegraphics[width=\textwidth]{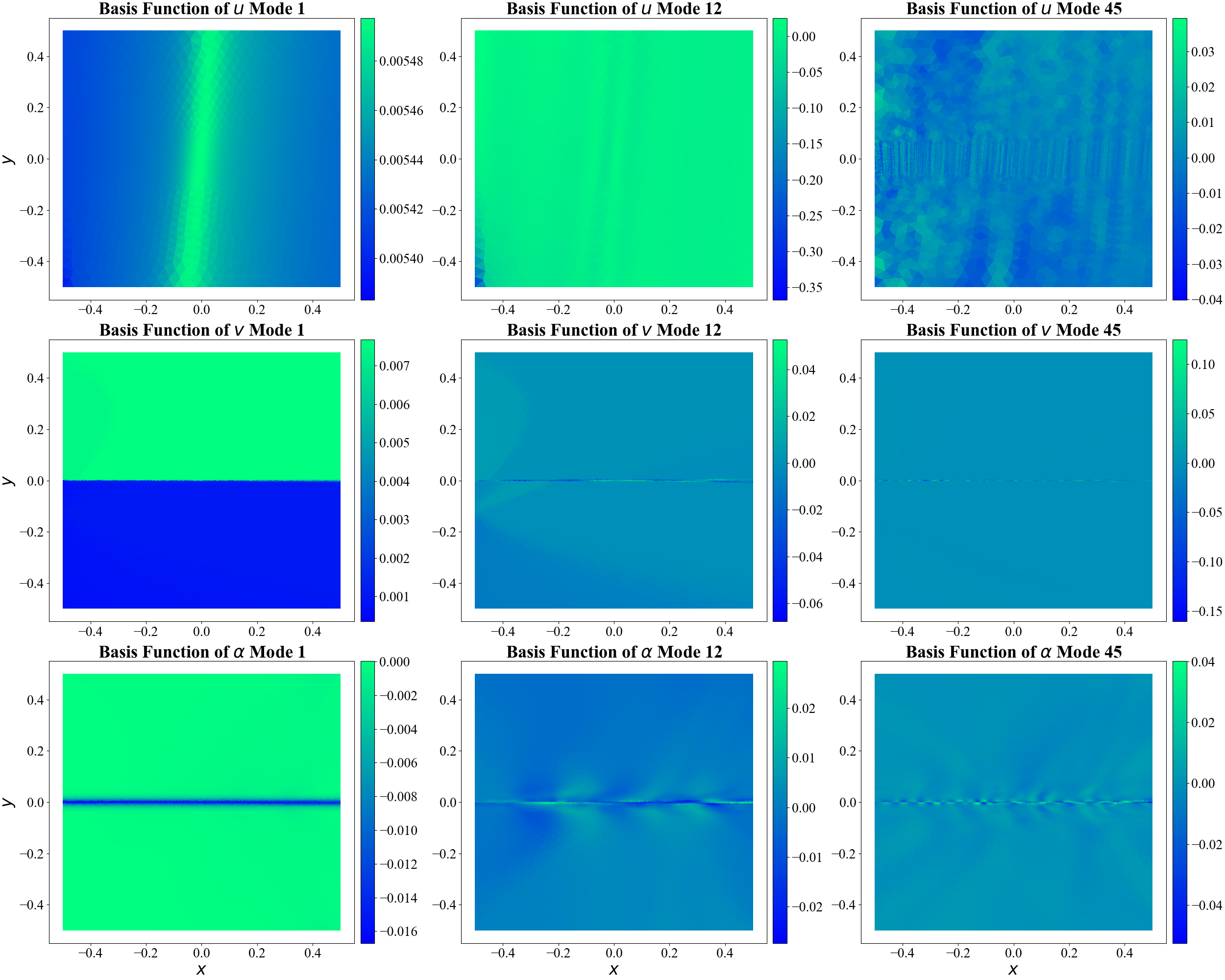}
        \caption{\textbf{Case 2}:  Tensile loading.}
        \label{fig:basis_function_tensil}
    \end{subfigure}
    \hfill
    \begin{subfigure}[b]{0.495\textwidth}
        \centering
        \includegraphics[width=\textwidth]{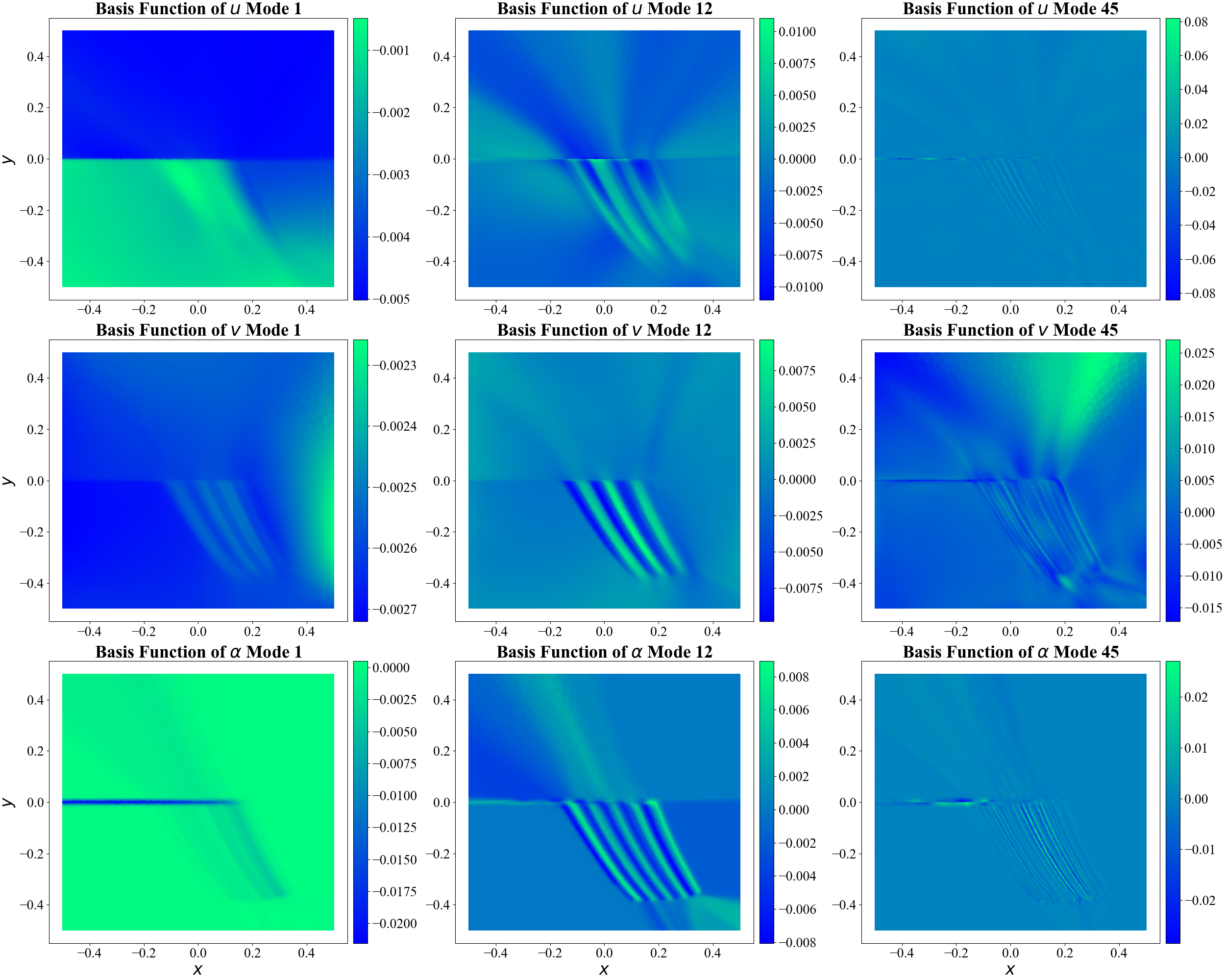}
        \caption{\textbf{Case 3}: Shear loading.}
        \label{fig:basis_function_shear}
    \end{subfigure}
        \hfill
    \begin{subfigure}[b]{0.495\textwidth}
        \centering
        \includegraphics[width=\textwidth]{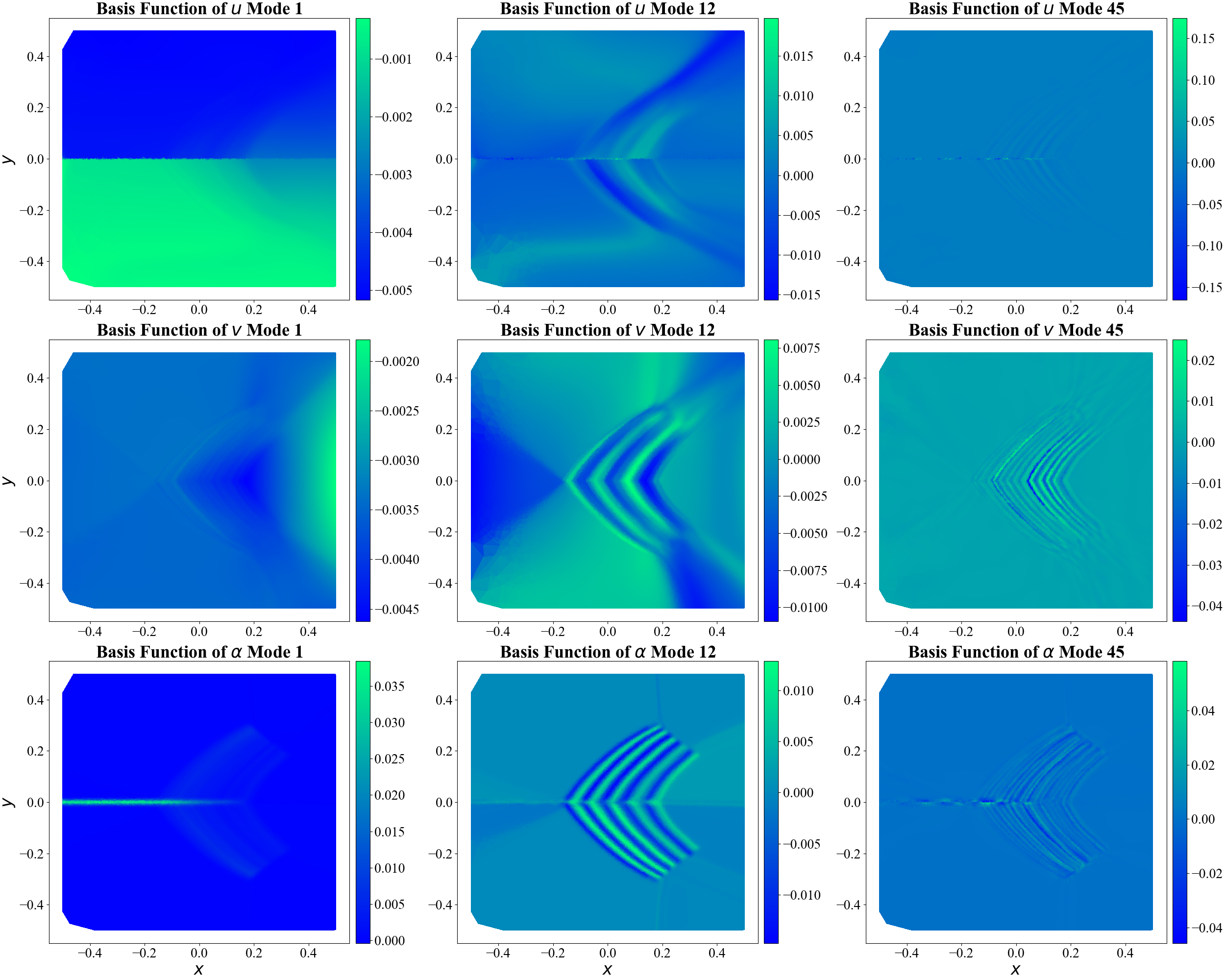}
        \caption{\textbf{Case 4}: Branching.}
        \label{fig:basis_function_branch}
    \end{subfigure}
\caption{Orthogonal basis functions, obtained from the output of the trunk network, corresponding to multiple modes of damage \(\alpha\) and displacement field components \(u\) and \(v\) for (a) \textbf{Case 2}, (b) \textbf{Case 3}, and (c) \textbf{Case 4}. In each plot, the first row corresponds to basis functions for \(u\), the second row corresponds to \(v\), and the third row corresponds to \(\alpha\).}
    \label{fig:Basis_function_value}
\end{figure}

We plot four basis functions for the 1D case in Figure~\ref{fig:basis_functions1D}. Remarkably, the first two modes qualitatively resemble the localized displacement and phase fields. Figure \ref{fig:Basis_function_value} compares the orthogonal basis functions for Mode 1, Mode 12, and Mode 45 for a SEN specimen under (a) tensile and (b) shear loading, as well as (c) branching. Mode 1 captures lower-frequency oscillations with smoother transitions, indicating its ability to model broad, global trends in crack propagation. 
These lower modes reflect large-scale deformations in the material and the gradual progression of damage as the crack propagates. In contrast, the higher modes exhibit higher-frequency oscillations with more localized variations. Higher modes capture finer details, such as localized deformation and sharp changes in damage around the crack tip. The increased oscillatory behavior observed in Mode 45 highlights the finer-scale responses in both displacement and damage near the crack tip.

\section{Prediction of fracture with physics-informed two-step DeepONet}\label{sec:Physics informed-Two-stepDeepont}

In this section, we discuss the performance of the physics-informed two-step DeepONet for the cases of crack propagation in SEN specimens subjected to tensile loading (\textbf{Case 2}) and shear loading (\textbf{Case 3}). Following the damage models used in data generation, we use the AT2 damage model for the tensile loading case and the AT1 damage model for the shear loading case. Again, the input to the trunk network is spatial coordinates, while the branch network processes the phase field associated with the initial notches. It is worth noting that physics-informed DeepONet is trained using only 10 samples, each with a different initial crack size, and tested on 5 additional samples.

\begin{figure}[!tbh]
    \centering
    \includegraphics[width=0.79\textwidth]{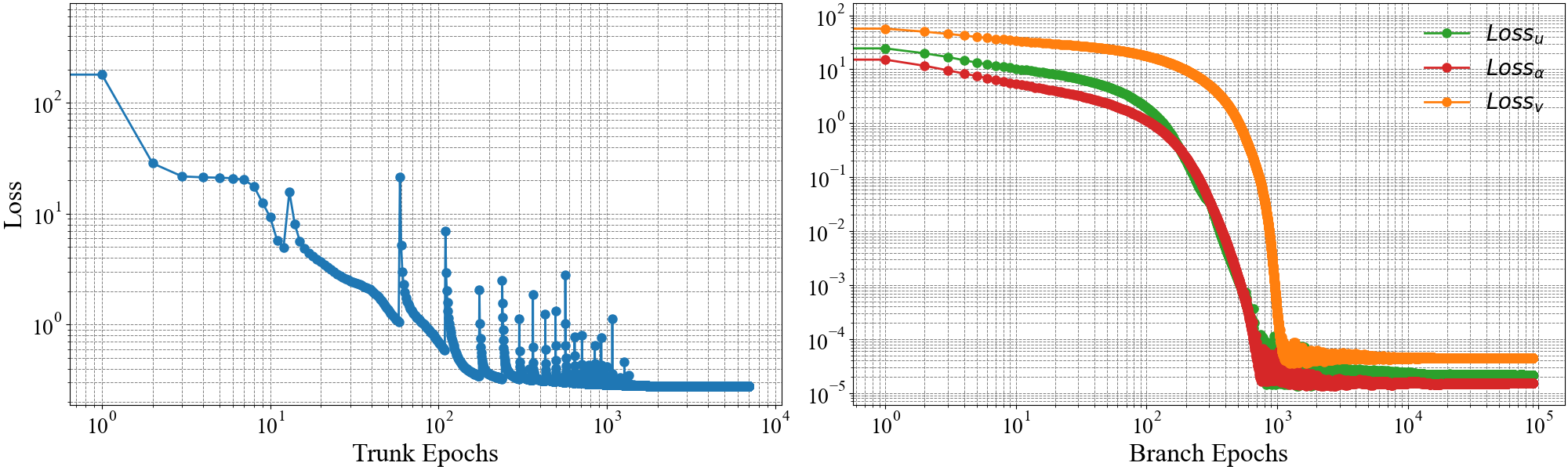}
\caption{\textbf{Physics-Informed Two-Step DeepONet (Case 2)}: The loss functions for the trunk network and branch network illustrate the evolution of the losses during training for predicting the displacement field components \( u \), \( v \), and the damage field \( \alpha \). The loss functions are balanced using \( \lambda_{\alpha} \), \( \lambda_{u} \), and \( \lambda_{v} \).}
\label{fig:loss_weights_comparison}
\end{figure}

\begin{figure}[!tbh]
    \centering
    \begin{subfigure}{0.49\textwidth}
        \centering
        \includegraphics[width=\textwidth]{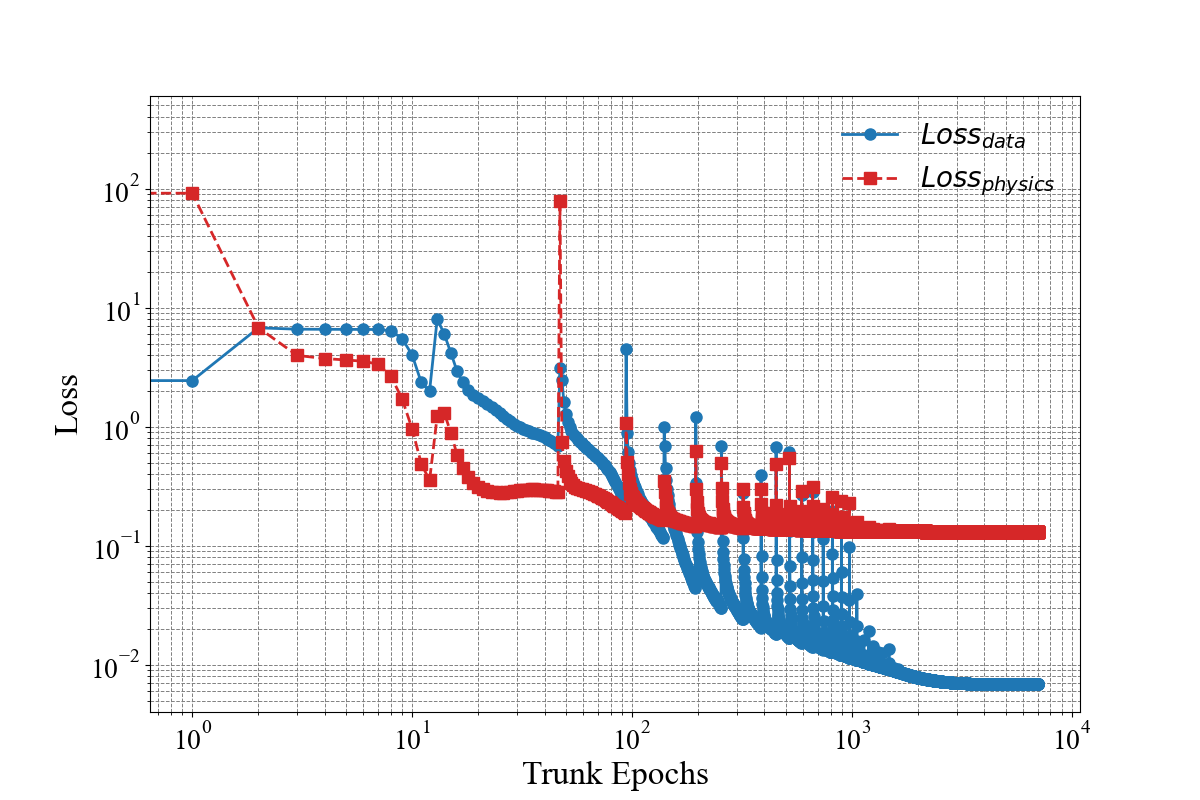}
        \caption{Loss function from trunk net}
        \label{fig:loss_function}
    \end{subfigure}
    \hfill
    \begin{subfigure}{0.49\textwidth}
        \centering
        \includegraphics[width=\textwidth]{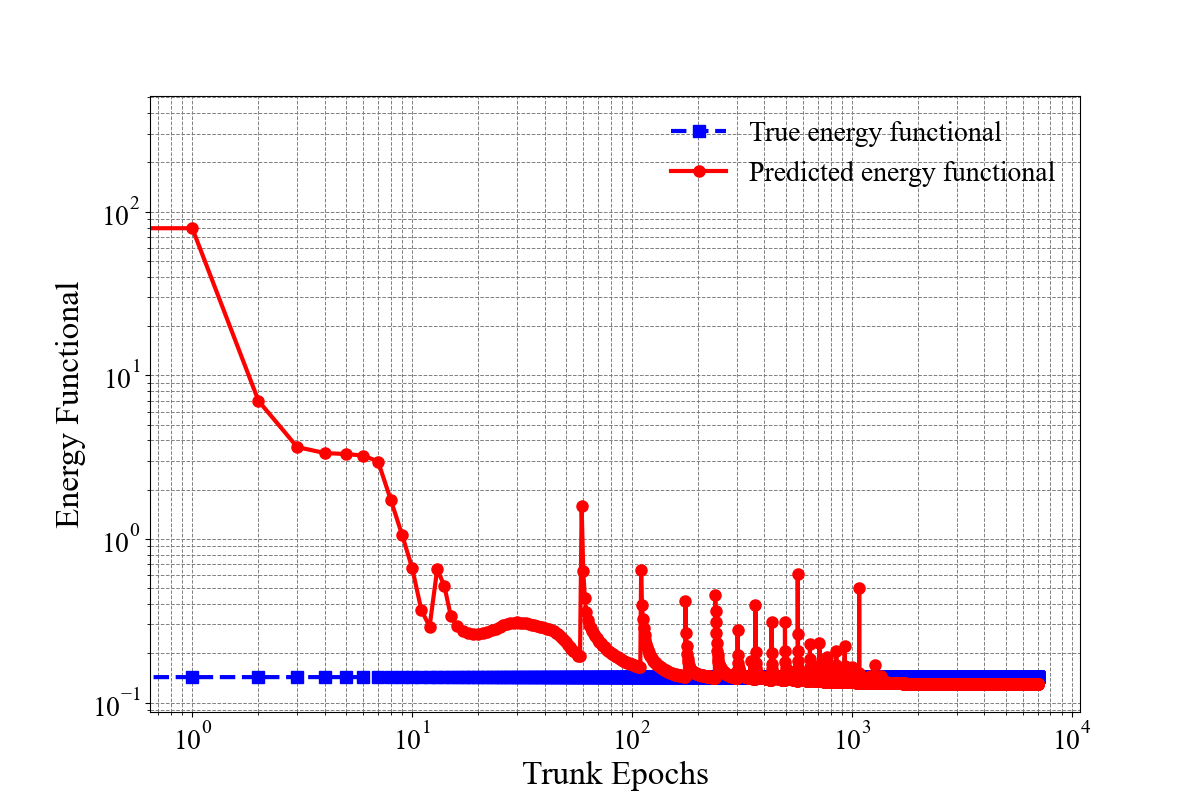}
        \caption{Energy functional}
        \label{fig:energy_function}
    \end{subfigure}
    \caption{\textbf{Physics-informed two-step DeepONet (Case 2).} A comparison of \( \texttt{loss}_{data} \) and \( \texttt{loss}_{physics} \) for the trunk net, as well as the energy functional generated by the FE simulations and the predicted energy functional in the physics-informed two-step DeepONet.}
    \label{fig:loss_comparison}
\end{figure}

Figures~\ref{fig:loss_weights_comparison} and~\ref{fig:loss_comparison} compare the behavior of the loss functions in the trunk and branch networks within the physics-informed two-step DeepONet. The plots in Figure~\ref{fig:loss_weights_comparison} illustrate how the trainable weights for \( \texttt{loss}_{u} \) help balance the contributions of the losses throughout training, as \( \texttt{loss}_{u} \) is relatively small compared to the other losses.
Figures~\ref{fig:loss_comparison}(a) shows that the initial value of \( \texttt{loss}_{\text{physics}} \), representing the energy functional term, starts at \(10^2\), significantly higher than \( \texttt{loss}_{data} \). Throughout the training process, \( \texttt{loss}_{\text{physics}} \) gradually stabilizes around \(10^{-1}\), while \( \texttt{loss}_{data} \) continues to decrease and eventually approaches \(10^{-2}\). As evident in the figure, the red line representing \( \texttt{loss}_{\text{physics}} \) changes very little after $\sim 10^3$ epochs, indicating that it has stabilized. At this point, the model shifts its focus exclusively to minimizing \( \texttt{loss}_{data} \), ensuring that the smaller loss components are effectively managed, contributing to balanced and efficient training. Moreover, Figure~\ref{fig:loss_comparison}(b) compares the predicted energy functional during training with the  energy functional obtained from the FE simulations.
Moreover, Table~\ref{Table:errors_physics} presents the MAEs for the displacement components \( u \), \( v \), and damage field \(\alpha\). 
The table shows that predicting displacement \( u \) is challenging due to its small values, leading to an error of \( 10^{-3} \) in the physics-informed two-step approach. Since \( u \) was used unscaled in the energy-based loss, the network struggled with predicting these small values.

\begin{figure}[!tbh]
    \centering
    \begin{subfigure}[b]{0.495\textwidth}
        \centering
        \includegraphics[width=\textwidth]{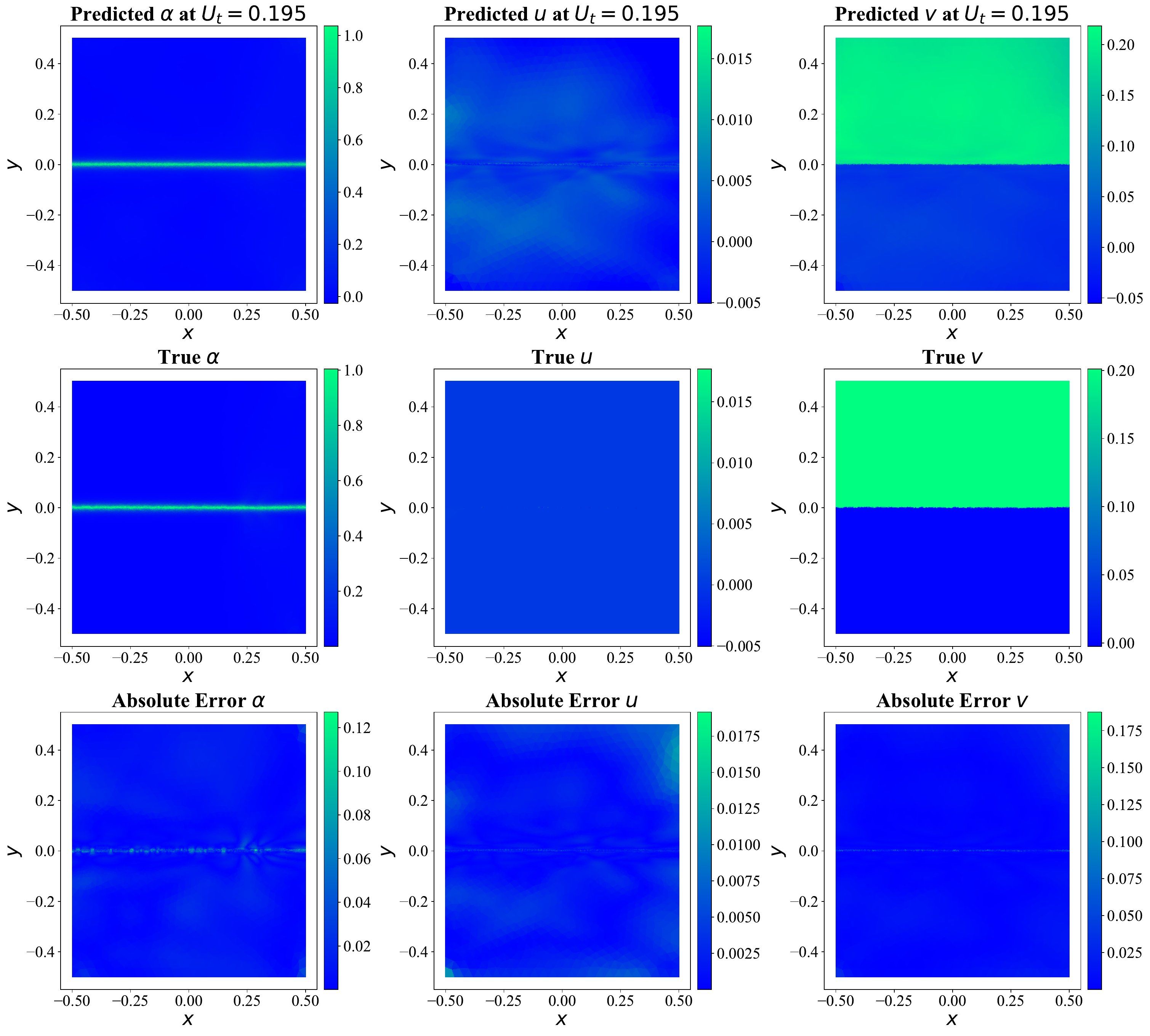}
        \caption{\textbf{Case 2}: Tensile loading}
        \label{fig:physics_case1_tensile}
    \end{subfigure}
    \hfill
    \begin{subfigure}[b]{0.495\textwidth}
        \centering
        \includegraphics[width=\textwidth]{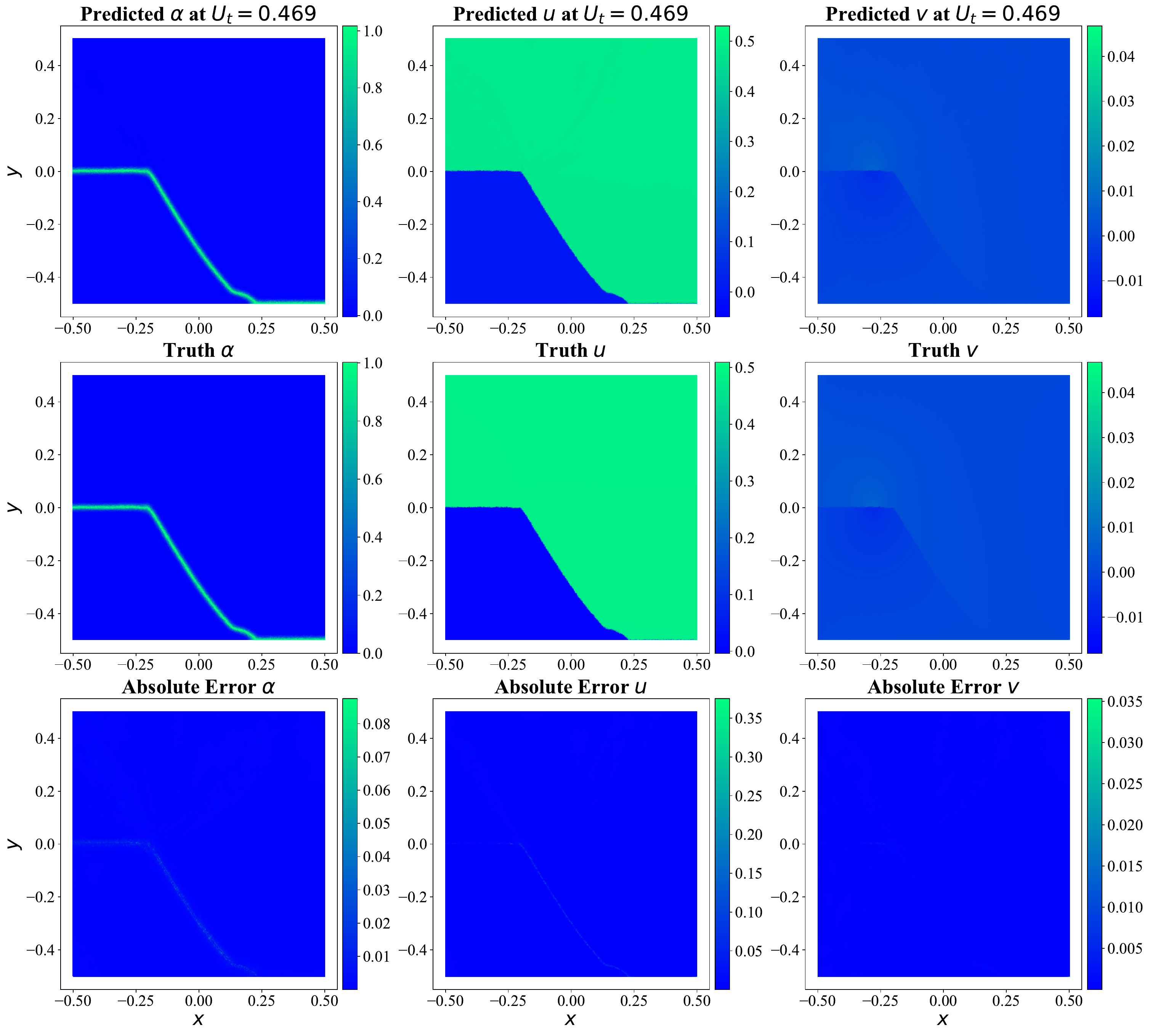}
        \caption{\textbf{Case 3}: Shear loading}
        \label{fig:physics_case2_SEN}
    \end{subfigure}
    \caption{\textbf{Physics-informed two-step DeepONet}: Comparison of the true and predicted damage \(\alpha\) and displacement field components \(u\) and \(v\) using the physics-informed two-step DeepONet. 
    The results are shown for two loading conditions: (a) \textbf{Case 2} - tensile loading at applied displacement value \(U_t = 0.195\), and (b) \textbf{Case 3} - shear loading at applied displacement value \(U_t = 0.469\). In each plot, the first column displays the damage field \(\alpha\), while the second and third columns show the displacement field components \(u\) and \(v\), respectively. The top two rows present the predicted values and the ground truth, and the third row shows the absolute error between the predictions and the true values.}
    \label{fig:prediction_physics_tensil_SEN}
\end{figure}

The plots in Figure~\ref{fig:prediction_physics_tensil_SEN} illustrate the results of the physics-informed two-step DeepONet, comparing the predicted and true values for \(\alpha\), \(u\) and \(v\) for a test sample. The figures correspond to (a) tensile loading in \textbf{Case 2} at a displacement of \(U_t = 0.195\), and (b) shear loading in \textbf{Case 3} at \(U_t = 0.469\). The displacement and phase fields away from the crack are accurately predicted by the networks in both cases. However, in the tensile loading case, predicted fields close to the crack exhibit large errors. In the shear loading case, fields close to the crack are also accurately predicted.

\begin{table}[h!]
    \centering
    \begin{tabular}{|c|c|c|c|}
        \hline
        \textbf{Case} & \textbf{MAE  \(u\)} & \textbf{MAE  \(v\)} & \textbf{MAE \(\alpha\)} \\
        \hline
        Tensile (Case 2) & 1.09e-03 & 5.36e-03 & 1.28e-02 \\
        \hline
        Shear (Case 3) & 6.58e-04 & 6.44e-05 & 6.60e-04 \\
        \hline
    \end{tabular}
\caption{\textbf{Physics-informed two-step DeepONet}: MAE for  \( u \), \( v \), and \(\alpha\) for \textbf{Case 2} and \textbf{Case 3}. 
}
    \label{Table:errors_physics}
\end{table}

\begin{table}[ht]
\centering
\small
\renewcommand{\arraystretch}{1.2} 
\setlength{\tabcolsep}{2pt} 
\begin{tabular}{|l|c|c|c|c|c|}
\hline
\textbf{Case} &  \textbf{ Trunk} & \textbf{Branch} & \textbf{Comp. Time (Trunk)} & \textbf{Comp. Time ($\alpha$)} & \textbf{Comp. Time (u/v)} \\ 

\hline 
\textbf{Case 2}   & 700000 epochs& 900000 epochs& 20029 s & 1361 s & 1365/1361 s\\                  
\hline
\textbf{Case 3}   & 500000 epochs& 700000 epochs& 98956 s & 2564 s & 2566/2576 s \\ 
\hline
\end{tabular}
\caption{\textbf{Physics-informed two-step DeepONet}: Training details for physics-informed two-step DeepONet. The sensor points listed in Table~\ref{table_examples} are used for training the trunk network, while the branch network is trained using 10 different initial sizes and tested on 5 samples.}
\label{Table_training_summary_Physics}
\end{table}

The MAEs using the 5 specimens reserved for testing are presented in Table~\ref{Table:errors_physics}. MAEs are larger for the fields in the tensile loading case compared to the shear loading case as also observed in Figure~\ref{fig:prediction_physics_tensil_SEN}. The value of MAEs for the three fields suggest that the network predictions are in close agreement with the FE data. Furthermore, Table~\ref{Table_training_summary_Physics} provides the training time for both the trunk and branch networks.

\section{Prediction of fracture with data-driven DeepOKAN}\label{sec:KAN}

In our DeepOKAN, the branch net comprises an MLP architecture
and $\texttt{tanh}$ as activation function. The trunk net employs a Chebyshev KAN structure~\cite{ss2024chebyshev}, utilizing learnable functions parameterized by Chebyshev polynomials of degree three, which is particularly well-suited to provide the 
basis for our operator learning task. We analyze the performance of DeepOKAN in \textbf{Case 1}, \textbf{Case 2}, and \textbf{Case 4}. 

\begin{figure}[!tbh]
    \centering
    \includegraphics[width=0.60\textwidth]{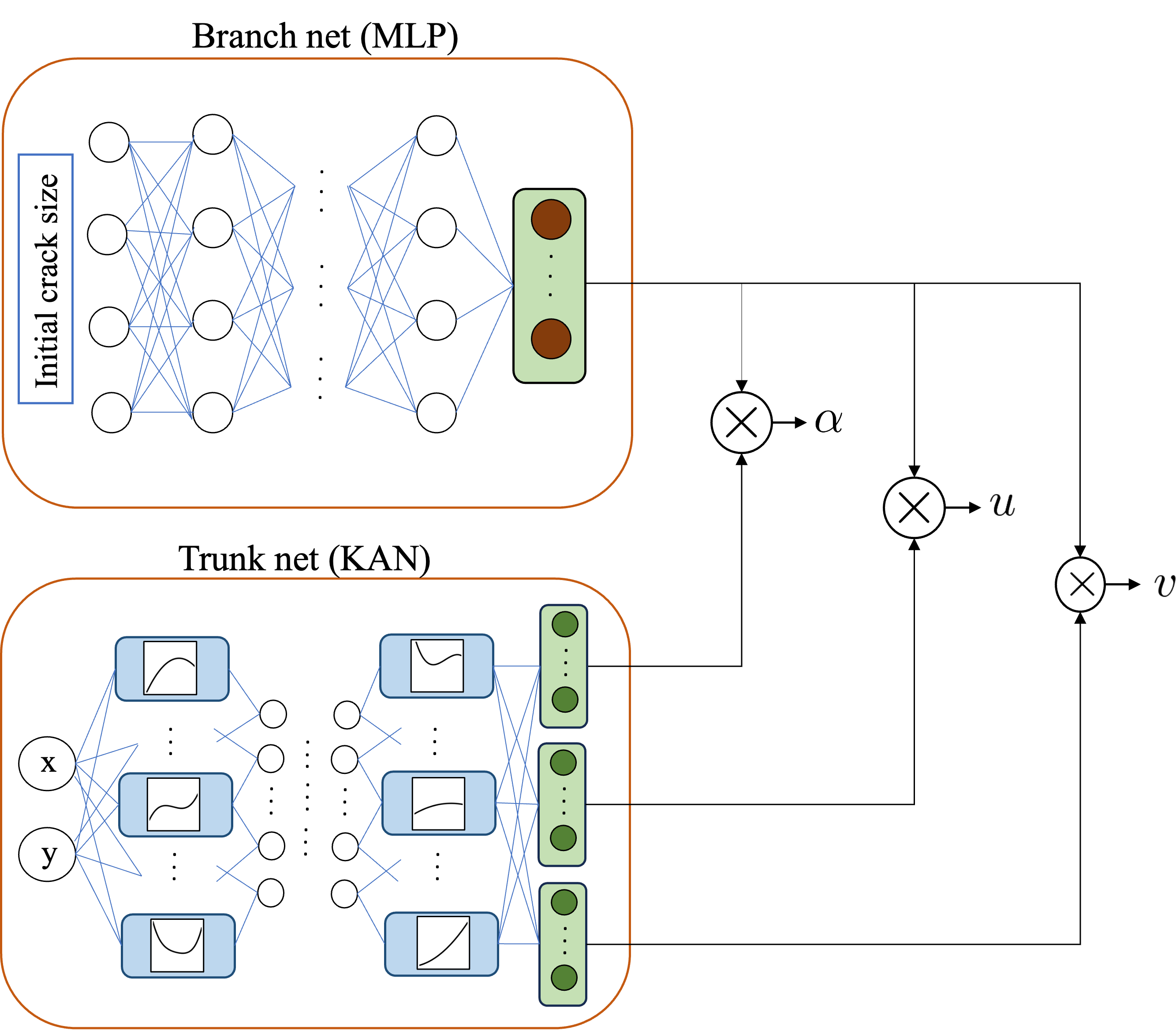}
\caption{\textbf{DeepOKAN}: Schematic of the DeepOKAN architecture comprising an MLP in the branch network and a KAN in the trunk network. 
The last layer of the KAN is split into three equal parts, and their inner product is performed with the outputs from the branch network, leading to three outputs (damage field \(\alpha\) and displacement field components \(u\) and \(v\)).}
    \label{fig:DeepOKAN_architecture}
\end{figure}

\begin{figure}[!tbh]
    \centering
    \begin{subfigure}[b]{0.995\textwidth}
        \centering
        \includegraphics[width=\textwidth]{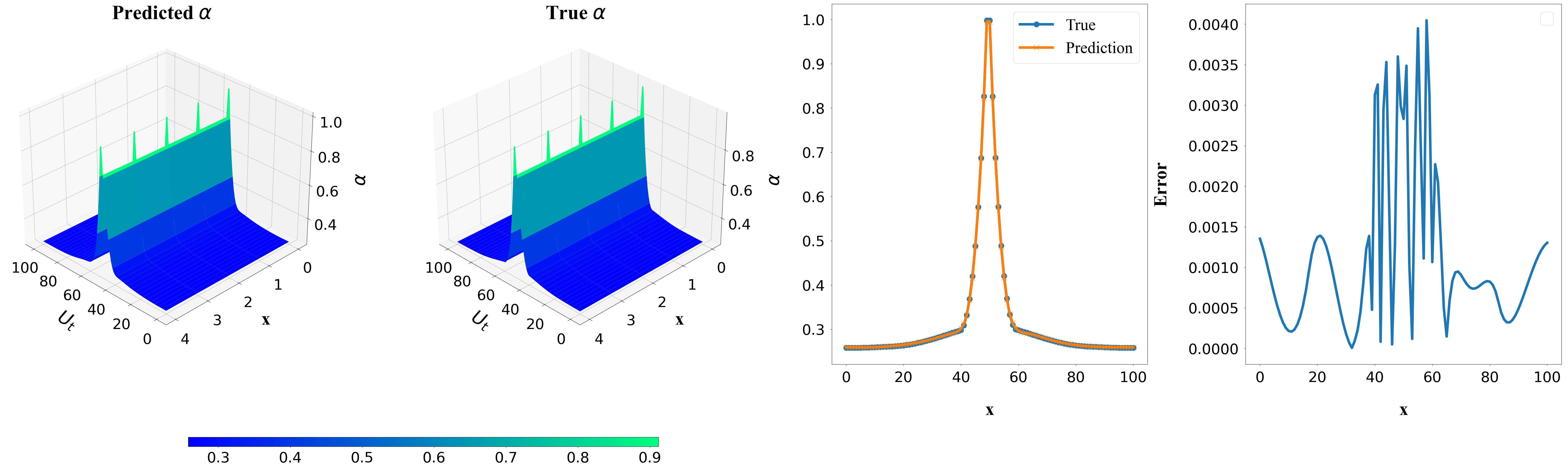}
        \caption{\textbf{Case 1}: Predicted and true damage \( \alpha \).}
        \label{fig:KAN_1D}
    \end{subfigure}
    \hfill
    \begin{subfigure}[b]{0.995\textwidth}
        \centering
        \includegraphics[width=\textwidth]{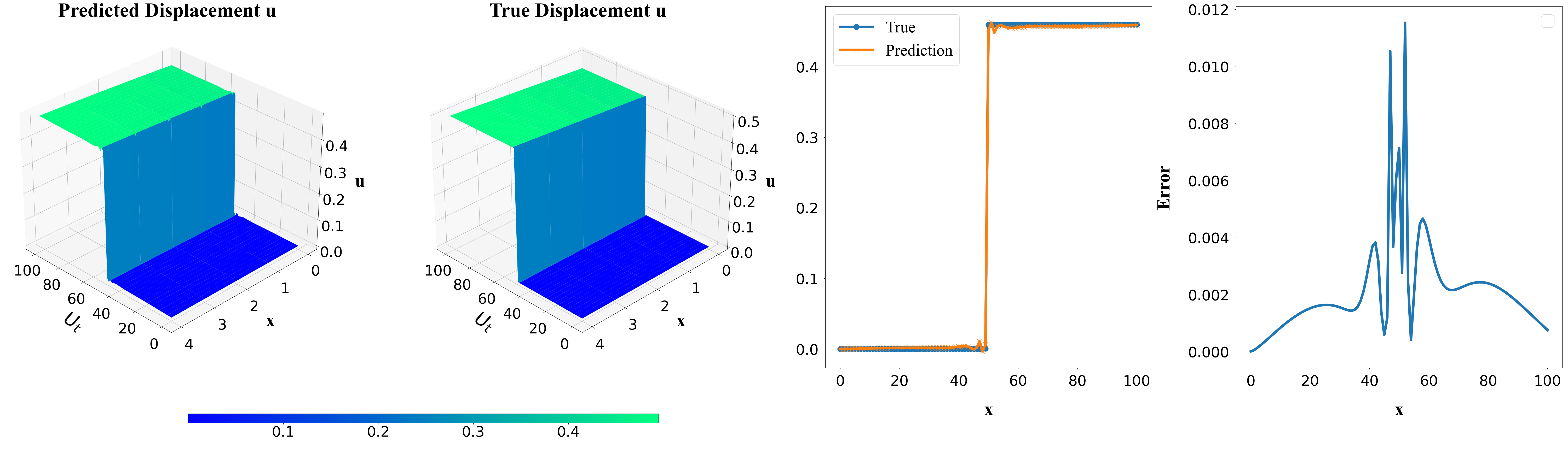}
        \caption{\textbf{Case 1}: Predicted and true displacement field \( u \).}
        \label{fig:KAN_1D}
    \end{subfigure}
    \caption{\textbf{DeepOKAN}: Comparison of predicted and true (a) damage \( \alpha \) and (b) displacement field \( u \), along with the absolute error between the predicted and true values. 
    } 
    \label{fig:DeepOKAN_1D}
\end{figure}

For \textbf{Case 1}, the branch network processes the applied displacement values \( U_t \) over 45 steps, while the trunk network handles the spatial coordinate of \( 101 \) nodes. Both the trunk and branch networks consist of 5 hidden layers with 50 neurons each, with the final layer also containing 50 neurons. The results of DeepOKAN for \textbf{Case 1} are reported in Figure~\ref{fig:DeepOKAN_1D}, where (a) shows the predicted and true damage \( \alpha \), and (b) displays the displacement field \( u \). The figures on the right illustrate the absolute error between the predicted and true values. The network predictions are largely in agreement with the FE data. However, the displacement field predicted by the network exhibits oscillations near the sharp change in the field resembling Gibbs phenomena.

For \textbf{Case 2} and \textbf{Case 4}, data from 45 randomly selected initial notch sizes are used for training, while the remaining 5 notch sizes are reserved for testing. The branch network processes the phase fields associated with the initial crack sizes, while the trunk network handles the spatial coordinate points.
Each subnetwork in DeepOKAN consists of 7 hidden layers, with 100 nodes per layer. The final layer of the branch network has 300 neurons, while the final layer of the trunk network contains 900 nodes. The outputs from each subnetwork are combined through an inner product to generate the final predictions, as shown in Figure~\ref{fig:DeepOKAN_architecture}. The best results are obtained by training the DeepOKAN to minimize the sum of the relative mean squared errors in \textbf{Case 2} and the sum of the mean squared errors in \textbf{Case 4}. This is different than for the previously analyzed DeepONets, for which we only minimized the sum of the mean squared errors. Overall, based on our experience, DeepOKAN is highly sensitive to hyperparameter tuning.

\begin{figure}[!tbh]
    \centering
    \begin{subfigure}[b]{0.495\textwidth}
        \centering
        \includegraphics[width=\textwidth]{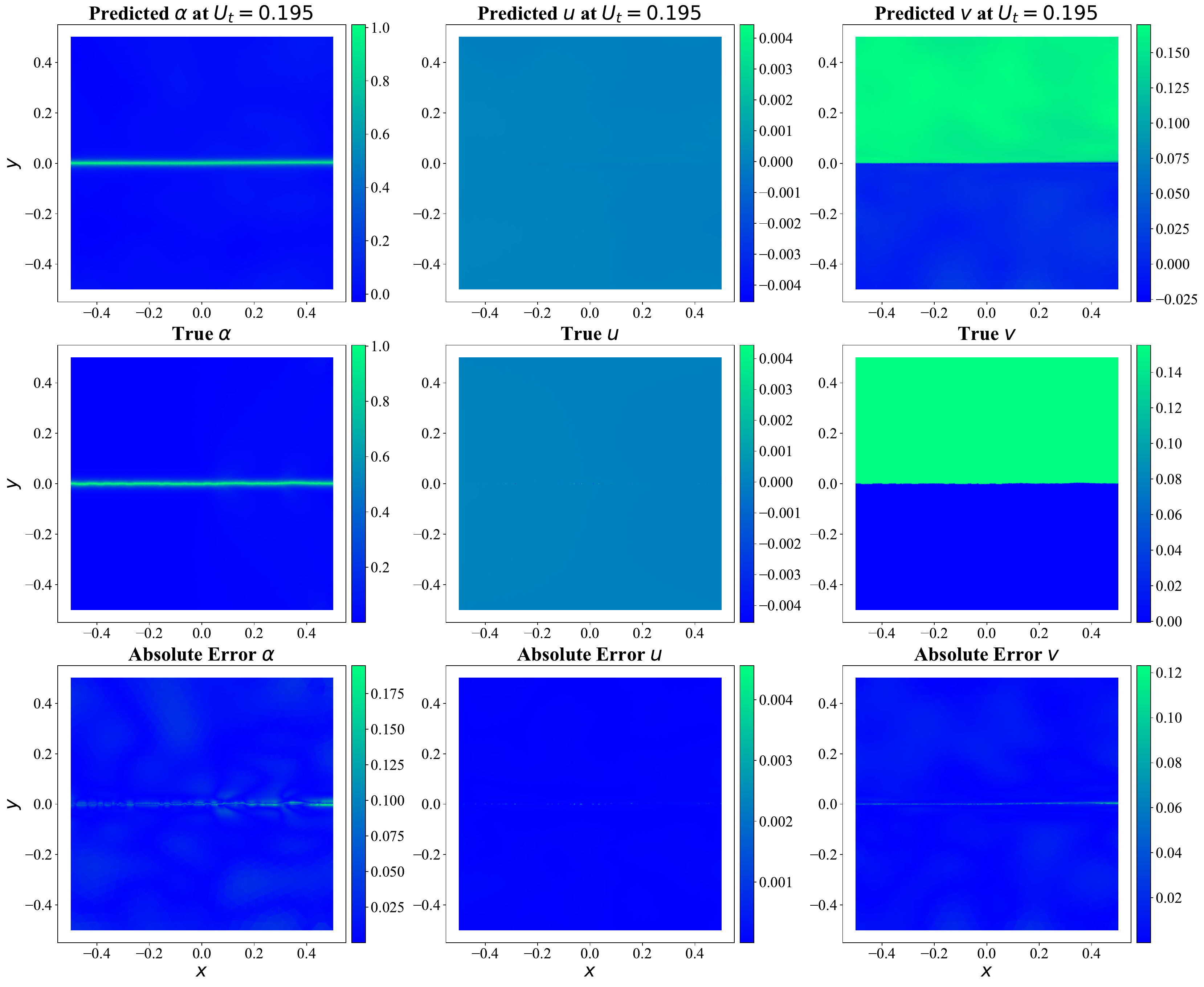}
        \caption{\textbf{Case 2}: Tensile Loading Conditions}
        \label{fig:KAN_tensil}
    \end{subfigure}
    \hfill
    \begin{subfigure}[b]{0.495\textwidth}
        \centering
        \includegraphics[width=\textwidth]{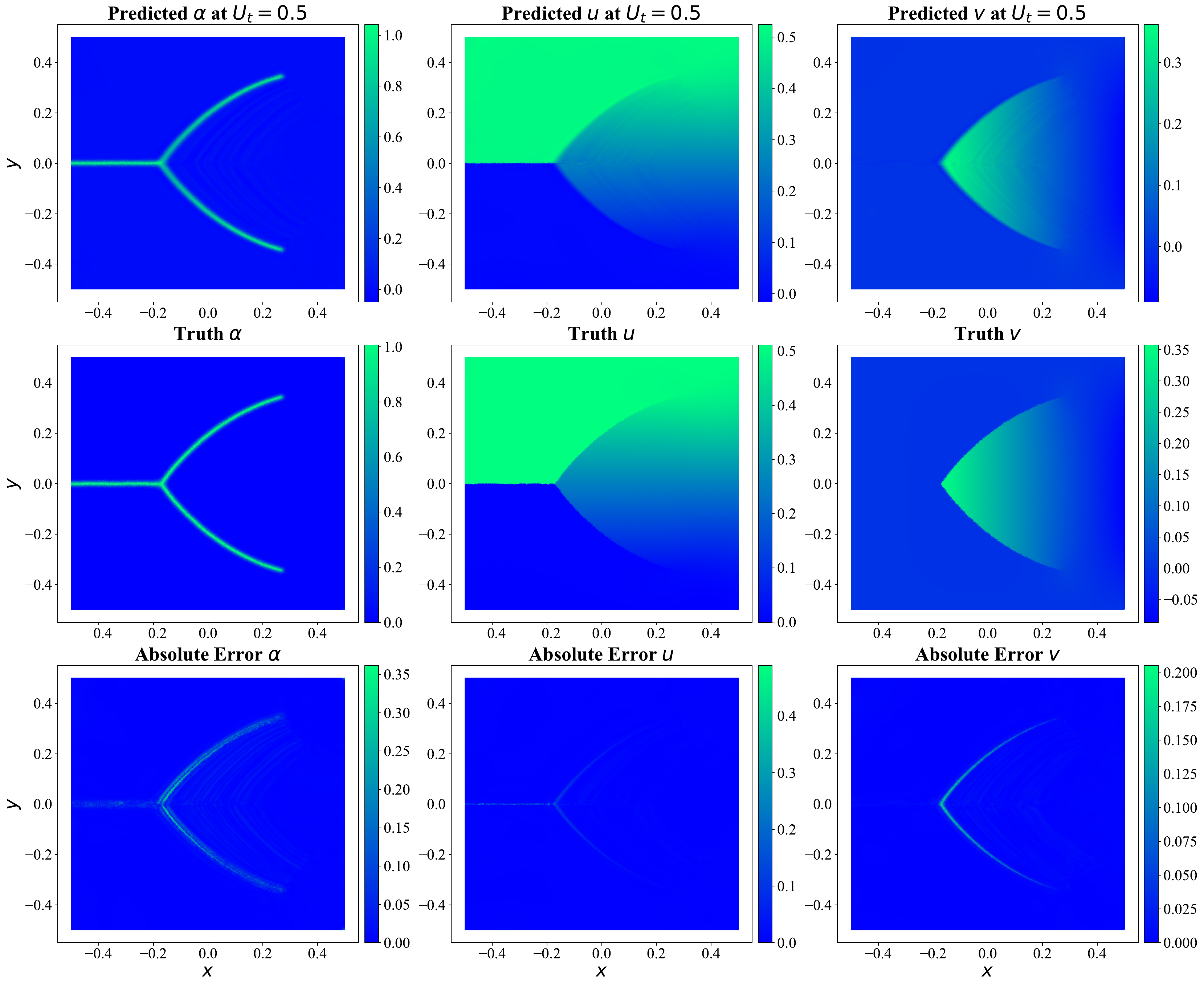}
        \caption{\textbf{Case 4}: Branching}
        \label{fig:KAN_branch}
    \end{subfigure}
    \caption{\textbf{DeepOKAN}: Comparison of true and predicted damage \(\alpha\) and displacement field components \(u\) and \(v\) using DeepOKAN. The results are presented for two different loading conditions: (a) \textbf{Case 2} - tensile loading at applied displacement value \(U_t = 0.195\), and (b) \textbf{Case 4} branching at applied displacement value \(U_t = 0.5\). 
     In each plot, the first column displays the damage field \(\alpha\), while the second and third columns show the displacement field components \(u\) and \(v\), respectively. The first two rows in each plot represent the predicted values and the ground truth, while the third row illustrates the corresponding absolute error between the predicted and true values.} 
    \label{fig:DeepOKAN_tensil}
\end{figure}

 Figure~\ref{fig:DeepOKAN_tensil} shows a comparison of the true and predicted damage and displacement fields using DeepOKAN for the applied displacement value \(U_t = 0.195\) and \(U_t = 0.5\). The predicted fields are in agreement with the FE data for both cases except near the crack where the error is localized. Error is larger in the branching case compared to the tensile loading case.
 
 The MAEs for crack propagation in an SEN specimen under tensile loading (\textbf{Case 2}) and shear with branching (\textbf{Case 4})—are presented in Table ~\ref{Table:errors_DeepOKAN}. The MAEs for the displacement field components \(u\), \(v\), and the phase field \(\alpha\) generally range from \(10^{-3}\) to \(10^{-2}\). The higher MAE for \(\alpha\) among the three network outputs indicates that the network finds it more challenging to approximate the phase field. Additionally, the MAEs for both the displacement and phase field are lower in \textbf{Case 2} compared to \textbf{Case 4}. This suggests that the operator mapping in \textbf{Case 4} is more difficult for DeepOKAN to learn than \textbf{Case 2}. Table~\ref{Table:errors_DeepOKAN_time} presents the training time and the number of epochs for each case.

\begin{table}[h!]
    \centering
    \begin{tabular}{|c|c|c|c|}
        \hline
        \textbf{Case} & \textbf{MAE  \(u\)} & \textbf{MAE  \(v\)} & \textbf{MAE  \(\alpha\)} \\
        \hline
        Tensile (Case 2) & 1.97e-03 & 1.82e-03 & 3.36e-02 \\
        \hline
         Branching (Case 4) & 3.74e-03 & 3.30e-03 & 9.73e-03 \\
        \hline
    \end{tabular}
\caption{\textbf{DeepOKAN}: mean absolute error for  \( u \), \( v \), and \(\alpha\) for \textbf{Case 2} and \textbf{Case 4}.}
\label{Table:errors_DeepOKAN}
\end{table}

\begin{table}[ht]
\centering
\small
\renewcommand{\arraystretch}{1.2} 
\setlength{\tabcolsep}{4pt} 
\begin{tabular}{|l|c|c|}
\hline
\textbf{Case} & \textbf{Epochs } & \textbf{Comp. Time} \\  
\hline
\textbf{Case 1} & 300000 & 1239 s \\                   
\hline 
\textbf{Case 2} & 100000 & 397 s \\                   
\hline 
\textbf{Case 4} & 500000 & 5916 s \\                
\hline
\end{tabular}
\caption{\textbf{DeepOKAN}: Training details for various data-driven cases using DeepOKAN architectures. In all cases, the branch network is trained with 45 samples and tested with 5 samples, while the number of points used in the trunk network for each case is provided in Table \ref{table_examples}.}
\label{Table:errors_DeepOKAN_time}.
\end{table}

\section{Comparison of the three methods}
\label{sec:Comparison}

As discussed above, we evaluated the performance of DeepONet for predicting crack nucleation and propagation using three different trunk networks. A comparison of the results obtained using the three different approaches for the main cases 1-4 is presented below.

\begin{itemize}
\item \textbf{Case 1}.
In this case, we utilized a two-step DeepONet trained on 45 applied displacement values to predict the damage and displacement components at the last 5 displacement steps. Compared to the other cases, we employed a smaller neural network architecture in both the trunk and branch networks due to the one-dimensional nature of the problem. A comparison between the two-step DeepONet and DeepOKAN indicates that DeepOKAN requires smaller networks compared to the two-step DeepONet.

\item \textbf{Case 2}. Predicting the displacement \( u \) was particularly challenging due to its small magnitude. In the data-driven two-step DeepONet, employing 45 samples for training, we addressed this imbalance by adjusting the weights of each loss component by normalizing their contributions relative to the total loss and applying modifications to maintain numerical stability. This ensured that each loss component contributed meaningfully to the optimization process. This approach proved effective, allowing us to predict displacement \( u \) accurately. In the physics-informed approach, where we incorporated physics into the loss functions and used only 10 samples, predicting displacement \( u \) remained challenging, as shown in Figure~\ref{fig:prediction_physics_tensil_SEN}. In DeepOKAN we encountered the same issue. To mitigate this, we multiplied by a constant during training and eventually divided by the same constant to achieve accurate predictions of the displacement \( u \). Tables~\ref{Table:errors_data_driven} and \ref{Table:errors_physics} indicate that the error in the data-driven approach was approximately \( 10^{-4} \), while in the physics-informed approach the error was about \( 10^{-3} \). For predicting the displacement field \( v \) and the damage field \( \alpha \), all three scenarios performed similarly.

\item \textbf{Case 3}. This case was less challenging than \textbf{Case 2}. The two-step DeepONet performed effectively in both the data-driven and physics-informed approaches, achieving comparable results for predicting the damage field \( \alpha \) and the displacement field components \( u \) and \( v \). As shown in Tables~\ref{Table:errors_data_driven} and \ref{Table:errors_physics}, the errors were approximately \( 10^{-4} \), demonstrating consistent accuracy across  different methods.

\item \textbf{Case 4}. In this case, both the two-step DeepONet and DeepOKAN demonstrated similar performance when predicting the damage field \( \alpha \) and the displacement field components \( u \) and \( v \). As shown in Tables~\ref{Table:errors_data_driven} and \ref{Table:errors_DeepOKAN}, the errors were around \( 10^{-3} \), indicating consistent accuracy across these methods. This result highlights that predicting \textbf{Case 4} is more challenging than \textbf{Case 3}. The physics-informed approach, which requires predictions of the damage field \( \alpha \) and displacement field components \( u \) and \( v \) at all nodes, is computationally intensive. In contrast, the data-driven approach allows the selection of training samples from a subset of nodes, eliminating the need to use all data nodes for training, thereby reducing the computational cost.
\end{itemize}

In general, the results demonstrate the effectiveness of the two-step DeepONet in predicting crack propagation. In a data-driven setting, the two-step DeepONet performs well across all examples, showcasing its ability to accurately capture discontinuities. However, training the trunk and branch networks separately can be computationally expensive. While incorporating the energy functional into the loss functions reduces the number of training samples required, the physics-informed DeepONet is computationally expensive because it necessitates training the network at all nodes and minimizing the energy of the system. The DeepOKAN also performs effectively, and the network structures used for training the two-step DeepONet and DeepOKAN are highly compatible. The DeepOKAN achieves comparable performance to the two-step DeepONet while using a smaller network architecture. However, the DeepOKAN appears more sensitive to hyperparameter tuning. Considering the outcomes of other scenarios, the data-driven two-step DeepONet and the presented DeepOKAN method are recommended for balancing efficiency and accuracy. 

\section{Summary}
\label{sec:Summary}

We studied the application of three variants of DeepONets in modeling crack nucleation and propagation with varying boundary conditions or varying initial notch lengths in specimens as inputs to the networks.
Initially, we trained two-step DeepONets in a data-driven approach using 45 samples with varying initial notch sizes.
We then transitioned to a physics-informed framework, reducing the sample size to 10 and minimizing the energy function directly within the trunk network, ensuring physically consistent and robust predictions.
Our results demonstrate that the two-step DeepONet effectively captures complex fracture behaviors, including crack propagation, kinking and branching, with high accuracy. In the final phase, we explored the use of KAN within the DeepONet framework (DeepOKAN). Based on our results, KAN shows potential as an alternative architecture for operator learning in fracture mechanics, effectively capturing the sharply varying damage and displacement fields  in complex fracture scenarios. 
Future work will focus on the investigation of how these three DeepONet variants scale to more complex three-dimensional fracture scenarios.

\section*{Declaration of competing interest}
The authors declare that they have no known competing financial interests or personal relationships that could have appeared to influence the work reported in this paper.

\section*{Acknowledgments}
This research was primarily supported as part of the AIM for Composites, an Energy Frontier Research Center funded by the U.S. Department of Energy (DOE), Office of Science, Basic Energy Sciences (BES), under Award \#DE-SC0023389 (computational studies, data analysis). Additional funding was provided by GPU Cluster for Neural PDEs and Neural Operators  to Support MURI Research and Beyond, under Award \#FA9550-23-1-0671.
In addition, M.M and L.D.L acknowledge funding from the Swiss National Science Foundation through Grant \#200021-219407 ‘Phase-field modeling of fracture and fatigue: from rigorous theory to fast predictive simulations’.
Computing facilities were provided by the Center for Computation and Visualization, Brown University.

\appendix
\section{Adaptive weighing of loss terms}\label{sec:Adaptive weighing}

A dynamic normalization strategy is employed for adjusting loss weights based on the relative magnitudes of individual loss components. For instance, if a particular loss component, such as $\texttt{loss}_u$, becomes small, its corresponding weight $\lambda_u$ increases, thereby directing more focus to that component during optimization. Conversely, if a particular loss component is very large, its weight is reduced, diminishing its relative importance. A small constant $\epsilon$ is added to avoid division by zero when any individual loss component is very close to zero. The weights adjust in proportion to the relative magnitude of each loss term, ensuring that no single term is completely neglected or dominates excessively. 
By normalizing each weight using the \(\texttt{total}_{\text{loss}}\), the approach ties the relative importance of each loss term to the overall objective, with the weights explicitly normalized to sum to $1$.

\section{Training data generation}\label{sec:Tdg}
We employ FE simulations to generate the training data. First, we discretize the domain with elements of size \( l/5\) around the notch and in the regions where the crack is expected to propagate. Away from these regions, the element sizes increase up to \( 4l\). For the 1D problem, a constant element size of \( l/5\) is employed together with linear shape functions and one Gauss point per element. For the 2D problems, we employ quadrilateral elements with bilinear shape functions and two Gauss points per parametric direction. Furthermore, we model the initial notches by prescribing \(\alpha=1\) at the location of the notch and solving a so called recovery problem (i.e. the minimization of the fracture energy). To enforce the irreversibility of the phase field, we set the irreversibility threshold \(\gamma_{irr}=10^{-3}\). We employ the staggered solution algorithm together with the Newton–Raphson method to iteratively converge to the solution~\cite{gerasimov2019penalization}. An error tolerance on the residual norm of \(10^{-4}\) is set for the staggered scheme and
\(10^{-6}\) for the Newton–Raphson method. Moreover, a maximum of 500 iterations is set for the Newton–Raphson method and
1000 iterations for the staggered scheme. We employ the AT2 model for the 1D bar problem as well as the problem of tensile loading of SEN samples, and the AT1 model for the problem of shear loading of SEN samples (both kinking and branching cases). We perform the FE computations using the GRIPHFiTH code available at \\
https://gitlab.ethz.ch/compmech/GRIPHFiTH.

While FE discretization is distinct for each problem with a different notch size, the sensor points of the trunk network are fixed for each case. They are chosen to be the nodes of a discretization of the domain using triangular elements with element size \(l/5\) in the regions of the domain encompassing all the notches and crack paths present in the training data. Away from these regions, the element sizes increase up to \(4l\). The displacement and phase fields obtained from the FE simulations are mapped to the sensor points by interpolation.

\section{Energy computation}\label{sec:Energy computation}
To compute the energy using the expression in~\eqref{eq:2} from the displacement and phase fields, we need to compute the gradients of the fields and evaluate the integral. To this end, we discretize the domain  like in the FE method such that the nodes are the sensor points of the trunk network and the network yields field values at these points. Field values and their gradients in an element are computed like in the FE method as follows
\begin{align}
	&{\bm u}^h({\bm x})=\sum_{i=1}^n N^{i}\left(\boldsymbol{x}\right)\boldsymbol{u}^{i}, \qquad &\alpha^h({\bm x})=\sum_{i=1}^n N^{i}\left(\boldsymbol{x}\right)\alpha^{i},\nonumber \\
	&\nabla{\bm u}^h({\bm x})=\sum_{i=1}^n \boldsymbol{u}^{i}\otimes\nabla N^{i}\left(\boldsymbol{x}\right), \qquad &\nabla\alpha^h({\bm x})=\sum_{i=1}^n \alpha^{i}\,\nabla N^{i}\left(\boldsymbol{x}\right),
\end{align}
where the superscript $h$ denotes the FE approximation of the fields. ${N}^i({\bm x})$ is the shape function associated with node $i$ of the discretization with $i=1...n$ and $n$ as the number of nodes, and $\boldsymbol{u}^{i}$ and $\alpha^{i}$ are the values of $\boldsymbol{u}$ and $\alpha$ at the same node, respectively. We employ triangular elements for discretization and Gauss quadrature with one Gauss point to evaluate the integral.

\section{Stress field }\label{app:stress_fields}
This appendix presents the visualized stress components before crack propagation starts for \textbf{Case 2}, SEN under tensile loading. The stress fields plotted include $\sigma_{xx}$, $\sigma_{yy}$, $\sigma_{xy}$, and the von Mises stress $\sigma_{vM}$.

\begin{figure}[htbp]
    \centering
    \includegraphics[width=0.99\textwidth]{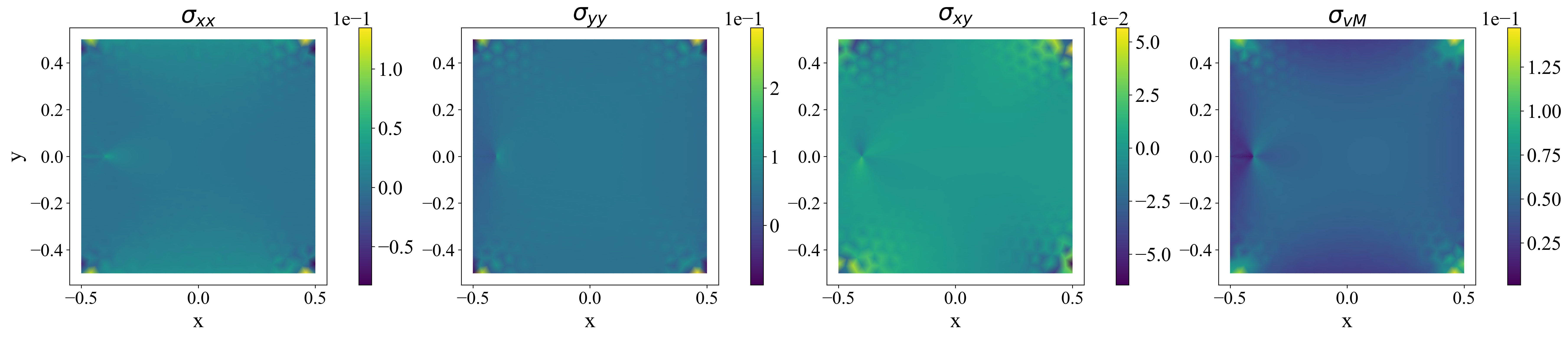}
    \caption{Visualization of the stress components $\sigma_{xx}$, $\sigma_{yy}$, $\sigma_{xy}$, and the von Mises stress $\sigma_{vM}$ for \textbf{Case 2} at an applied displacement value of \(U_t = 0.041\).}
    \label{fig:stress_components_vM}
\end{figure}

\bibliography{sample}

\begin{thebibliography}{10}
\expandafter\ifx\csname url\endcsname\relax
  \def\url#1{\texttt{#1}}\fi
\expandafter\ifx\csname urlprefix\endcsname\relax\def\urlprefix{URL }\fi
\expandafter\ifx\csname href\endcsname\relax
  \def\href#1#2{#2} \def\path#1{#1}\fi

\bibitem{griffith1921vi}
A.~A. Griffith, The phenomena of rupture and flow in solids, Philosophical Transactions of the Royal Society of London. Series A, containing papers of a mathematical or physical character (1921) 163--198.

\bibitem{francfort1998revisiting}
G.~A. Francfort, J.-J. Marigo, Revisiting brittle fracture as an energy minimization problem, Journal of the Mechanics and Physics of Solids (1998) 1319--1342.

\bibitem{bourdin2000numerical}
B.~Bourdin, G.~A. Francfort, J.-J. Marigo, Numerical experiments in revisited brittle fracture, Journal of the Mechanics and Physics of Solids (2000) 797--826.

\bibitem{bourdin2008variational}
B.~Bourdin, G.~A. Francfort, J.-J. Marigo, The variational approach to fracture, Journal of Elasticity (2008) 5--148.

\bibitem{ambati2015review}
M.~Ambati, T.~Gerasimov, L.~De~Lorenzis, A review on phase-field models of brittle fracture and a new fast hybrid formulation, Computational Mechanics (2015) 383--405.

\bibitem{bourdin2014morphogenesis}
B.~Bourdin, J.-J. Marigo, C.~Maurini, P.~Sicsic, Morphogenesis and propagation of complex cracks induced by thermal shocks, Physical review letters 112~(1) (2014) 014301.

\bibitem{luo2023phase}
C.~Luo, L.~Sanavia, L.~De~Lorenzis, Phase-field modeling of drying-induced cracks: Choice of coupling and study of homogeneous and localized damage, Computer Methods in Applied Mechanics and Engineering 410 (2023) 115962.

\bibitem{heider2021review}
Y.~Heider, A review on phase-field modeling of hydraulic fracturing, Engineering Fracture Mechanics 253 (2021) 107881.

\bibitem{alessi2018comparison}
R.~Alessi, M.~Ambati, T.~Gerasimov, S.~Vidoli, L.~De~Lorenzis, Comparison of phase-field models of fracture coupled with plasticity, Advances in computational plasticity: A book in honour of D. Roger J. Owen (2018) 1--21.

\bibitem{lorenzis2020numerical}
L.~De~Lorenzis, T.~Gerasimov, Numerical implementation of phase-field models of brittle fracture, in: Modeling in Engineering Using Innovative Numerical Methods for Solids and Fluids, Springer, 2020, pp. 75--101.

\bibitem{fuhg2024review}
J.~Fuhg, G.~Padmanabha, N.~Bouklas, W.~Sun, N.~Vlassis, M.~Flaschel, P.~Carrara, L.~De~Lorenzis, A review on data-driven constitutive laws for solids, Archives of Computational Methods in Engineering (2024).

\bibitem{raissi2019physics}
M.~Raissi, P.~Perdikaris, G.~E. Karniadakis, Physics-informed neural networks: A deep learning framework for solving forward and inverse problems involving nonlinear partial differential equations, Journal of Computational physics (2019) 686--707.

\bibitem{haghighat2022physics}
E.~Haghighat, D.~Amini, R.~Juanes, Physics-informed neural network simulation of multiphase poroelasticity using stress-split sequential training, Computer Methods in Applied Mechanics and Engineering (2022) 115141.

\bibitem{rasht2022physics}
M.~Rasht-Behesht, C.~Huber, K.~Shukla, G.~E. Karniadakis, Physics-informed neural networks (pinns) for wave propagation and full waveform inversions, Journal of Geophysical Research: Solid Earth (2022) e2021JB023120.

\bibitem{kiyani2024characterization}
E.~Kiyani, M.~Kooshkbaghi, K.~Shukla, R.~B. Koneru, Z.~Li, L.~Bravo, A.~Ghoshal, G.~E. Karniadakis, M.~Karttunen, Characterization of partial wetting by cmas droplets using multiphase many-body dissipative particle dynamics and data-driven discovery based on pinns, Journal of Fluid Mechanics (2024) A7.

\bibitem{kiyani2023framework}
E.~Kiyani, K.~Shukla, G.~E. Karniadakis, M.~Karttunen, A framework based on symbolic regression coupled with extended physics-informed neural networks for gray-box learning of equations of motion from data, Computer Methods in Applied Mechanics and Engineering (2023) 116258.

\bibitem{goswami2020transfer}
S.~Goswami, C.~Anitescu, S.~Chakraborty, T.~Rabczuk, Transfer learning enhanced physics informed neural network for phase-field modeling of fracture, Theoretical and Applied Fracture Mechanics 106 (2020) 102447.

\bibitem{manav2024phase}
M.~Manav, R.~Molinaro, S.~Mishra, L.~De~Lorenzis, Phase-field modeling of fracture with physics-informed deep learning, Computer Methods in Applied Mechanics and Engineering (2024) 117104.

\bibitem{zheng2022physics}
B.~Zheng, T.~Li, H.~Qi, L.~Gao, X.~Liu, L.~Yuan, Physics-informed machine learning model for computational fracture of quasi-brittle materials without labelled data, International Journal of Mechanical Sciences (2022) 107282.

\bibitem{chen2024crack}
Z.~Chen, Y.~Dai, Y.~Liu, Crack propagation simulation and overload fatigue life prediction via enhanced physics-informed neural networks, International Journal of Fatigue 186 (2024) 108382.

\bibitem{lu2019deeponet}
L.~Lu, P.~Jin, G.~Pang, Z.~Zhang, G.~E. Karniadakis, Learning nonlinear operators via {DeepONet} based on the universal approximation theorem of operators, Nature Machine Intelligence (2021) 218--229.

\bibitem{shih2024transformers}
B.~Shih, A.~Peyvan, Z.~Zhang, G.~E. Karniadakis, Transformers as neural operators for solutions of differential equations with finite regularity, arXiv preprint arXiv:2405.19166 (2024).

\bibitem{kovachki2023neural}
N.~Kovachki, Z.~Li, B.~Liu, K.~Azizzadenesheli, K.~Bhattacharya, A.~Stuart, A.~Anandkumar, Neural operator: Learning maps between function spaces with applications to pdes, Journal of Machine Learning Research (2023) 1--97.

\bibitem{shukla2024deep}
K.~Shukla, V.~Oommen, A.~Peyvan, M.~Penwarden, N.~Plewacki, L.~Bravo, A.~Ghoshal, R.~M. Kirby, G.~E. Karniadakis, Deep neural operators as accurate surrogates for shape optimization, Engineering Applications of Artificial Intelligence (2024) 107615.

\bibitem{oommen2024integrating}
V.~Oommen, A.~Bora, Z.~Zhang, G.~E. Karniadakis, Integrating neural operators with diffusion models improves spectral representation in turbulence modeling, arXiv preprint arXiv:2409.08477 (2024).

\bibitem{CiCP-35-1194}
L.~Liu, K.~Nath, W.~Cai, A causality-deeponet for causal responses of linear dynamical systems, Communications in Computational Physics (2024) 1194--1228.

\bibitem{wang2021learning}
S.~Wang, H.~Wang, P.~Perdikaris, Learning the solution operator of parametric partial differential equations with physics-informed deeponets, Science Advances (2021) eabi8605.

\bibitem{lee2024training}
S.~Lee, Y.~Shin, On the training and generalization of deep operator networks, SIAM Journal on Scientific Computing (2024) C273--C296.

\bibitem{peyvan2024riemannonets}
A.~Peyvan, V.~Oommen, A.~D. Jagtap, G.~E. Karniadakis, Riemannonets: Interpretable neural operators for riemann problems, Computer Methods in Applied Mechanics and Engineering (2024) 116996.

\bibitem{li2023phase}
W.~Li, M.~Z. Bazant, J.~Zhu, Phase-field deeponet: Physics-informed deep operator neural network for fast simulations of pattern formation governed by gradient flows of free-energy functionals, Computer Methods in Applied Mechanics and Engineering (2023) 116299.

\bibitem{GOSWAMI2022114587}
S.~Goswami, M.~Yin, Y.~Yu, G.~E. Karniadakis, A physics-informed variational {DeepONet} for predicting crack path in quasi-brittle materials, Computer Methods in Applied Mechanics and Engineering (2022) 114587.

\bibitem{shukla2024comprehensive}
K.~Shukla, J.~D. Toscano, Z.~Wang, Z.~Zou, G.~E. Karniadakis, A comprehensive and fair comparison between mlp and kan representations for differential equations and operator networks, arXiv preprint arXiv:2406.02917 (2024).

\bibitem{liu2024kan}
Z.~Liu, Y.~Wang, S.~Vaidya, F.~Ruehle, J.~Halverson, M.~Solja{\v{c}}i{\'c}, T.~Y. Hou, M.~Tegmark, Kan: {Kolmogorov-Arnold Networks}, arXiv preprint arXiv:2404.19756 (2024).

\bibitem{sprecher2002space}
D.~A. Sprecher, S.~Draghici, Space-filling curves and {Kolmogorov} superposition-based neural networks, Neural Networks (2002) 57--67.

\bibitem{koppen2002training}
M.~K{\"o}ppen, On the training of a kolmogorov network, in: Artificial Neural Networks—ICANN 2002: International Conference, Madrid, Spain, August 28--30, 2002, pp. 474--479.

\bibitem{pham2011}
K.~Pham, H.~Amor, J.~Marigo, C.~Maurini, Gradient damage models and their use to approximate brittle fracture, International Journal of Damage Mechanics 20 (2011) 618–652.

\bibitem{amor2009regularized}
H.~Amor, J.-J. Marigo, C.~Maurini, Regularized formulation of the variational brittle fracture with unilateral contact: Numerical experiments, Journal of the Mechanics and Physics of Solids (2009) 1209--1229.

\bibitem{freddi2009variational}
F.~Freddi, G.~Royer-Carfagni, Variational models for cleavage and shear fractures, in: Proceedings of the XIX AIMETA Symposium, 2009, pp. 715--716.

\bibitem{miehe2010thermodynamically}
C.~Miehe, F.~Welschinger, M.~Hofacker, Thermodynamically consistent phase-field models of fracture: Variational principles and multi-field fe implementations, International Journal for Numerical Methods in Engineering (2010) 1273--1311.

\bibitem{vicentini2024energy}
F.~Vicentini, C.~Zolesi, P.~Carrara, C.~Maurini, L.~De~Lorenzis, On the energy decomposition in variational phase-field models for brittle fracture under multi-axial stress states, International Journal of Fracture (2024) 1--27.

\bibitem{gerasimov2019penalization}
T.~Gerasimov, L.~De~Lorenzis, On penalization in variational phase-field models of brittle fracture, Computer Methods in Applied Mechanics and Engineering (2019) 990--1026.

\bibitem{wang2024expressiveness}
Y.~Wang, J.~W. Siegel, Z.~Liu, T.~Y. Hou, On the expressiveness and spectral bias of {KANs}, arXiv preprint arXiv:2410.01803 (2024).

\bibitem{kendall2018multi}
A.~Kendall, Y.~Gal, R.~Cipolla, Multi-task learning using uncertainty to weigh losses for scene geometry and semantics, in: Proceedings of the IEEE conference on computer vision and pattern recognition, 2018, pp. 7482--7491.

\bibitem{li2022revisiting}
W.~Li, C.~Zhang, C.~Wang, H.~Guan, D.~Tao, Revisiting pinns: Generative adversarial physics-informed neural networks and point-weighting method, arXiv preprint arXiv:2205.08754 (2022).

\bibitem{chen2024self}
W.~Chen, A.~A. Howard, P.~Stinis, Self-adaptive weights based on balanced residual decay rate for physics-informed neural networks and deep operator networks, arXiv preprint arXiv:2407.01613 (2024).

\bibitem{kingma2014adam}
D.~P. Kingma, Adam: A method for stochastic optimization, arXiv preprint arXiv:1412.6980 (2014).

\bibitem{ss2024chebyshev}
S.~SS, K.~AR, A.~KP, Chebyshev polynomial-based {Kolmogorov-Arnold Networks}: An efficient architecture for nonlinear function approximation, arXiv preprint arXiv:2405.07200 (2024).

\end{thebibliography}

\end{document}